%% file: main.tex
\documentclass[12pt,a4paper,oneside,onecolumn]{book}
\usepackage[a4paper, margin=1in]{geometry}
\usepackage{pdfpages}

\usepackage{libertine}
\usepackage{courier}
\usepackage[english]{babel}
\usepackage[longnamesfirst,square,numbers,comma,sort&compress]{natbib}
\usepackage{titlesec}
\usepackage{float}
\usepackage{graphicx}
\usepackage{subcaption}
\usepackage{amsmath}
\usepackage{mathpazo}
\usepackage{datetime}
\usepackage{setspace}
\usepackage{amsmath}
\usepackage{etoolbox}
\usepackage{rotating} 
\usepackage{array}    
\usepackage{soul}
\usepackage{xcolor}
\usepackage{longtable}
\usepackage{booktabs}
\usepackage{amssymb} 
\usepackage{multicol}
\usepackage{tabularx} 

\usepackage{hyperref}
\hypersetup{colorlinks=true, linkcolor=blue, urlcolor=blue, citecolor=blue}

\sethlcolor{yellow}

\makeatletter
\renewcommand{\tagform@}[1]{,\ \textnormal{\normalfont(#1)}}
\makeatother

\titleformat{\chapter}[hang]{\bf\huge}{\thechapter}{2pc}{}
\title{HUJI M.Sc.~Thesis Template}

\begin{document}

\onehalfspacing
\input{0.cover}

\frontmatter

\renewcommand{\baselinestretch}{1.5}

\input{1.abstract}
\input{2.acknowledgements}
\input{2.5.contribution}
\hypersetup{
    linkcolor=black, 
}
\tableofcontents
\listoffigures
\listoftables
\hypersetup{
    linkcolor=blue, 
}

\mainmatter

\input{3.introduction}
\input{4.theoretical_background}

\input{5.experimental_setup}
\input{6.analysis_and_results}

\input{7.conclusion}

\bibliography{thesis}

\end{document}

%% file: 0.cover.tex
\begin{titlepage}
    \begin{center}
        \vspace*{1cm}
        
        \includegraphics[width=0.2\textwidth, height=0.3\textwidth]{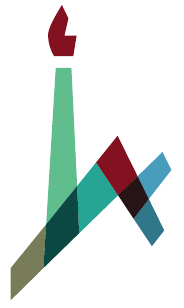}

        The Hebrew University of Jerusalem\\
        The Racah Institute of Physics
        
        \vspace{2cm}
        
        {\Large \textbf{\Large \( \beta \)-Decay Spectroscopy of \textsuperscript{23}Al}}
        
        \vspace{1cm}
        
        \textbf{Itay Goldberg}
        
        \vspace{1cm}
        
        Thesis submitted in partial fulfillment of the requirements\\for the Master of Sciences degree\\
        in Physics
        
        \vspace{1cm}
        
        Under the supervision of \textbf{Dr.~Moshe Friedman}
        
        \vfill
        
        \textbf{December 2024}
    \end{center}
\end{titlepage}

%% file: 1.abstract.tex
\chapter*{Abstract}

This research explores the \(\beta\) decay of the proton-rich nucleus \(^{23}\text{Al}\). The nucleus was generated at the National Superconducting Cyclotron Laboratory (NSCL) through projectile fragmentation, utilizing a primary beam of \(^{36}\text{Ar}\) ions directed at a \(^{9}\text{Be}\) target. Simultaneous measurements of proton emission and \(\gamma\) rays were conducted using the GADGET detection system. The decay paths were carefully analyzed through \(\beta\)-\(\gamma\), proton-\(\gamma\), and \(\gamma\)-\(\gamma\) coincidences, leading to the construction of a complete decay scheme for \(^{23}\text{Al}\). The absolute \(\beta\) branching ratios were determined, and log-ft values were calculated for transitions to \(^{23}\text{Mg}\) states. Additionally, proton branching ratios and the most precise half-life measurement of \(^{23}\text{Al}\) to date were obtained. The findings include the identification of 19 new \(\gamma\) rays and the discovery of a new \(\beta\)-delayed proton transition populating the third excited state of \(^{22}\text{Na}\).

%% file: 2.acknowledgements.tex
\chapter*{Acknowledgements}

I would like to express my sincere gratitude to all those who supported me throughout my Master's journey.

First and foremost, I extend my heartfelt thanks to my supervisor, Dr. Moshe Friedman, for his invaluable guidance and insight throughout my Master's journey. His expertise and thoughtful feedback have played a crucial role in shaping my research and this thesis. I am particularly grateful for his kindness and exceptional consideration of my needs, which fostered a supportive and nurturing environment that greatly enriched my academic experience.

I am also deeply appreciative of the collaborative group at Michigan State University (MSU). I would like to thank Professor Chris Wrede for sharing his expertise, valuable advice, and thought-provoking ideas. My gratitude goes to Dr. Lijie Sun for his thorough assistance and careful review of my work, as well as for his significant contributions to the SeGA efficiency calculations.

I would also like to specifically thank Mr. Aviv Bello for his technical support. Additionally, I am grateful to Ms. Shachar Ben Dor for her kind administrative assistance whenever I needed support; her help was greatly appreciated.

Lastly, I wish to thank my friends and family for their unwavering encouragement and belief in me throughout this academic journey. Your support has been a vital source of motivation and has consistently kept my spirits high.

%% file: 2.5.contribution.tex
\chapter*{Letter of Contribution}

The research and results presented in this thesis are the outcome of a collaborative effort. The thesis was written under the supervision and guidance of Dr. Moshe Friedman.

I wrote of the Introduction and Background chapters (Chapters 1 and 2). Additionally, I wrote the description of the experimental setup provided in Chapter 3, though I was not directly involved in constructing or performing the experiment itself.

Regarding the experimental work, my personal contributions are primarily reflected in the analysis presented in Chapter 4 (with the exception of Section \ref{sec:sim_eff}) and the discussion provided in Chapter 5.

This statement outlines the scope of my direct involvement in this collaborative work.

Sincerely,

Itay Goldberg

%% file: 3.introduction.tex
\chapter{Introduction}
\label{chap:intro}
\begin{figure}[h!]
    \centering
    \includegraphics[width=\textwidth]{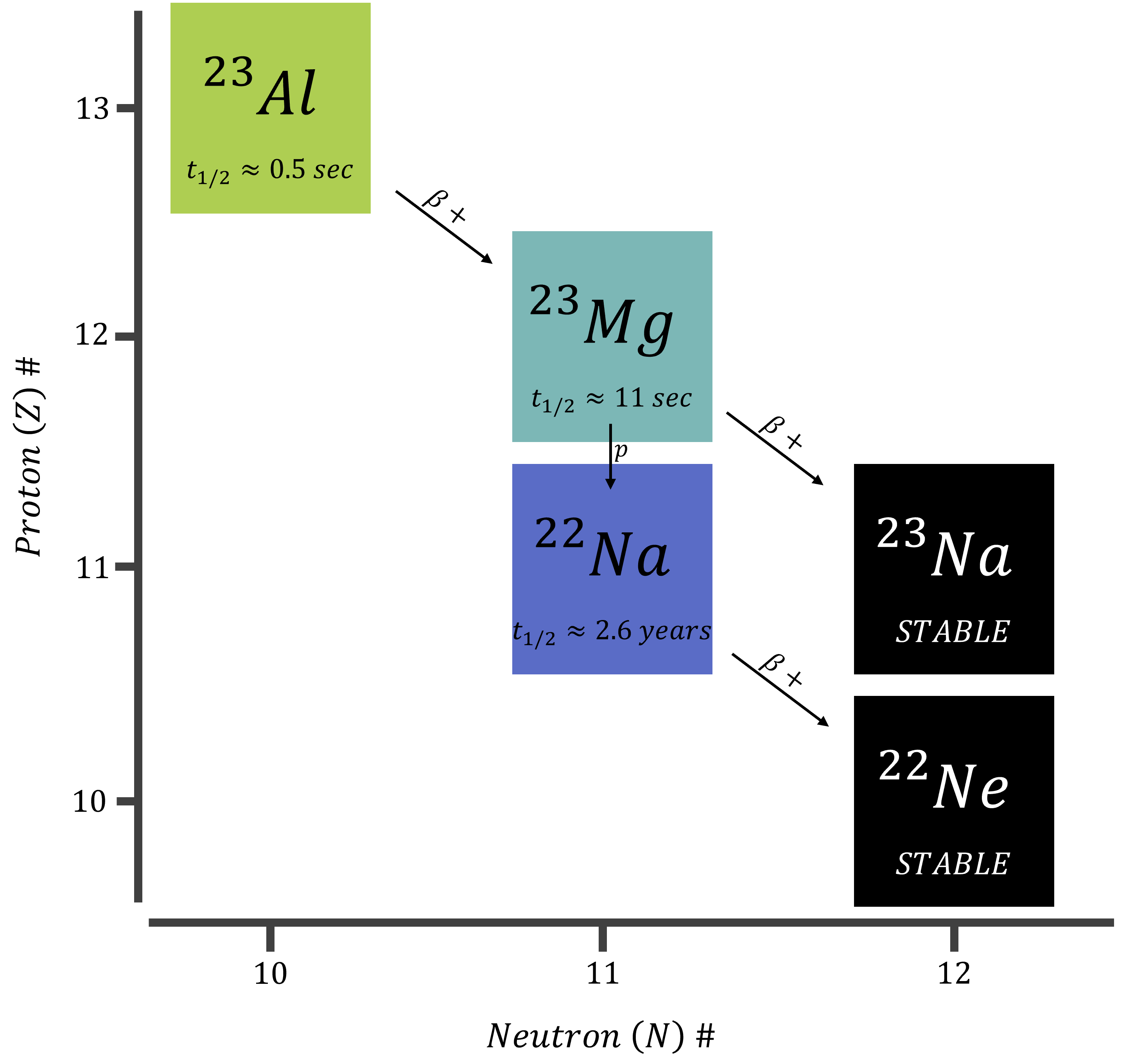}
    \caption{Decay paths of \textsuperscript{23}Al and its subsequent products.}
    \label{fig:isotopes}
\end{figure}

The study of nuclear characteristics of unstable isotopes is crucial for advancing our understanding of fundamental nuclear processes and is essential for integrating knowledge about the various isotopes that exist in nature and their interdependencies. In this research, the focus is placed on the isotope \textsuperscript{23}Al, which undergoes \( \beta \) decay into \textsuperscript{23}Mg. \textsuperscript{23}Mg can further emit a proton, decaying into \textsuperscript{22}Na, or decay via \( \beta \) decay into \textsuperscript{23}Na. The decay chain eventually stabilizes with \textsuperscript{23}Na and \textsuperscript{22}Ne, as shown in Figure~\ref{fig:isotopes}.

The motivation for this experiment was to measure the \( \gamma \) rays and protons following the initial \( \beta \) decay of \textsuperscript{23}Al. A significant aspect was to investigate a transition pertinent to \( \gamma \)-ray astronomy. The radionuclide \textsuperscript{22}Na is important in thermonuclear runaways in classical novae, as the \textsuperscript{22}Na(p, $\gamma$) \textsuperscript{23}Mg reaction serves as its primary destruction channel, affecting \textsuperscript{22}Na yields. Prior investigations have reported inconsistent results regarding resonance strengths, as highlighted in a Rapid Communication by Friedman et al. \cite{Friedman2020}. That Rapid Communication specifically addressed mainly the proton-related findings and is based on measurements from the same experiment as this thesis.

This thesis expands upon these initial findings by providing a comprehensive analysis of all recorded data from the experiment, including the populated energy levels of \textsuperscript{23}Mg, along with their corresponding \( \beta \) feeding branching ratios (BR) and Log-ft values, as well as the complete decay scheme of \textsuperscript{23}Al and its half-life. Firstly, it is critical to note that no comprehensive analysis of the \( \beta \)-delayed gamma emissions from \textsuperscript{23}Al has yet been conducted, hence ENSDF consider its decay scheme incomplete \cite{Basunia2021}; this work aims to fill that gap, enabling the construction of a full decay scheme for \textsuperscript{23}Al, whereas prior studies primarily focused on \( \beta \)-delayed protons. Secondly, the sophisticated detection system utilized in this study includes a low \( \beta \) noise proton detector along with \( \gamma \) detectors, enabling simultaneous measurement of \( \gamma \) rays and \( \beta \) particles. This capability, combined with high statistics, offers a unique opportunity to accurately assess the decay characteristics of \textsuperscript{23}Al and has the potential to reveal new transitions as well as refine previously known parameters.

In this work, \( \gamma \) and proton energies are meticulously sorted and traced back to their origins, along with their intensities, to construct a comprehensive decay scheme of \textsuperscript{23}Al.

\section{Previous Works}

The investigation of \textsuperscript{23}Al and its decay pathways has been explored through numerous experimental approaches. In the early 2000s, significant progress was achieved using various detection setups to capture the intricate details of \textsuperscript{23}Al decay. For instance, Saastamoinen et al. \cite{Saastamoinen2011}, Trache et al. \cite{Trache2012}, and Iacob et al. \cite{Iacob2006} studied \textsuperscript{23}Al produced via the \textsuperscript{1}H(\textsuperscript{24}Mg,\textsuperscript{23}Al) reaction. Specifically, Saastamoinen et al. focused on the \( \beta \)-delayed proton decay of \textsuperscript{23}Al, detecting protons down to an energy of 200 keV and determining corresponding branching ratios. Their findings highlighted a significant discrepancy with earlier studies, as no strong proton intensity from the decay of the isobaric analog state (IAS) at \( E_x = 7803 \) keV in \textsuperscript{23}Mg was observed. Instead, they identified a low-energy proton at \( E_{p,c.m.} = 206 \) keV, suggesting decay from a state \( 16 \) keV below the IAS. They measured both proton and \( \gamma \) branches from the decay of a state at \( E_x = 7787 \) keV, which is crucial for understanding the resonances involved in the radiative proton capture reaction \( \textsuperscript{22}Na(p,\gamma) \textsuperscript{23}Mg \).

Iacob et al. \cite{Iacob2006} conducted the first comprehensive study of the \( \beta \) decay of \textsuperscript{23}Al using pure samples obtained from the Momentum Achromat Recoil Separator (MARS). Their research allowed for \( \beta \) branching ratio and log ft value determinations for transitions to states in \textsuperscript{23}Mg, confirming the spin and parity of the \textsuperscript{23}Al ground state as \( J^\pi = 5/2^+ \). This result provides critical insights into nuclear astrophysics by ruling out significant increases in the radiative proton capture cross-section that would have resulted from different interpretations of spin and parity. Notably, their measurements focused on the IAS and the 16 keV lower state, both of which are significant for the \( \textsuperscript{22}Na(p,\gamma) \textsuperscript{23}Mg \) reaction in novae. Yongjun Zhai provided a more complete analysis of this experiment in his dissertation \cite{Zhai2007}, but the results were not published and not included in ENSDF evaluation \cite{Basunia2021}.

Kirsebom et al. \cite{Kirsebom2011} reported on a new measurement of the \( \beta \)-delayed proton spectrum of \textsuperscript{23}Al, achieving higher statistics compared to previous measurements. This substantial increase in statistics allowed them to identify new proton lines in the energy range of 1–2 MeV and perform a statistical analysis of the observed \( \beta \) strength. The enhanced statistical robustness of their results makes them particularly valuable for reference in this thesis.

Additional previous works are not specifically covered here as they do not directly contribute to the focus of this thesis. The existing literature highlights the critical need for further investigation of the decay characteristics and the emitted \(\gamma \) spectrum, areas where a comprehensive analysis has yet to be conducted. These aspects will be thoroughly explored in this thesis, contributing to a deeper understanding of the nuclear processes involved.

\section{Thesis outline}

The structure of this thesis is organized as follows: Chapter 2 provides the theoretical background, covering basic concepts radioactivity, decays, and details relevant to this research. Chapter 3 delves into the experimental setup, explaining how the experiment was conducted and how measurements were taken. The main part of this thesis is presented in Chapters 4 and 5. Chapter 4 comprehensively details the data analysis processes. Finally, Chapter 5 concludes the work by presenting the results and discussing the key findings.

%% file: 4.theoretical_background.tex
\chapter{Theoretical Background}
\label{chap:intro}

The theoretical background is primarily based on Krane's comprehensive work, \textit{Introductory Nuclear Physics} \cite{Krane1988}.

\section{The Radioactive Decay Law}

The radioactive decay law is characterized by the exponential growth or decay processes of these nuclei, governed by inherent probabilistic mechanisms. At its core, the law describes how the number of radioactive nuclei, $N(t)$, decreases over time according to the equation:

\begin{equation}
N(t) = N_0 e^{-\lambda t}
\end{equation}

where $N_0$ is the initial quantity of nuclei, $\lambda$ is the decay constant specific to a given isotope, and $t$ is time. The decay constant, $\lambda$, represents the probability per unit time that a nucleus will decay, providing a fundamental parameter for quantifying the rate at which radioactive material transforms.

A critical concept derived from the exponential decay law is the \textit{half-life} ($T_{1/2}$), which is linked to the decay constant through the relation:

\begin{equation}
T_{1/2} = \frac{\ln(2)}{\lambda}
\end{equation}

\section{\( \beta \) Decay Theory}

\( \beta \) decay is characterized by the emission of \( \beta \) particles, which can be either electrons or positrons, along with associated neutrinos or antineutrinos. The transformations associated with \( \beta \) decay are:

\begin{align*}
\text{Beta-minus decay:} \quad & n \rightarrow p + e^- + \bar{\nu}_e \\
\text{Beta-plus decay:} \quad & p \rightarrow n + e^+ + \nu_e \\
\text{Electron capture:} \quad & p + e^- \rightarrow n + \nu_e \\
\end{align*}

In \( \beta \) decay, the total number of nucleons remains constant; however, the transformation alters the isotope, changing one element into another. \( \beta \) particles have a considerable penetration ability, typically requiring a few millimeters of metal or dense material, like a sheet of aluminum, for effective shielding. The emitted \( \beta \) particles exhibit a continuous energy spectrum due to energy sharing among the emitted \( \beta \) and associated neutrino.

The decay is associated with a Q value, which represents the energy released during the decay. For a \( \beta \)+ decay it is calculated as:

\begin{align}
Q_{\beta^+} = \left( m_{N}\! \left( {}^{A}X \right) - m_{N}\! \left( {}^{A}X' \right) - m_{e^+} \right)c^2
\end{align}

where \(Q_{\beta^+}\) represents the released energy, \(m_{N}\! \left( {}^{A}X \right)\) denotes the mass of the initial nucleus \(X\) with mass number \(A\), \(m_{N}\! \left( {}^{A}X' \right)\) indicates the mass of the final nucleus \(X'\) with the same mass number, \(m_{e^+}\) refers to the mass of the emitted positron and \(c\) is the speed of light.

\subsection{Selection Rules}

In 1934, a significant theory describing beta decay was developed by Fermi. The fundamental aspects of this theory were derived from basic expressions for transition probabilities caused by weak interactions, compared to the interactions that form the initial and final states. Such transitions are described by a principle known as \textit{Fermi's Golden Rule}:

\begin{equation}
\lambda = \frac{2\pi}{\hbar} |V_{fi}|^2 \rho(E_f)
\label{golden_rule}
\end{equation}

where \(\lambda\) is the transition rate, \(V_{fi}\) represents the matrix element, computed as the integral over the interaction \(V\) between the initial and final states, and \(\rho(E_f)\) is the density of final states, expressed as \(\frac{dn}{dE_f}\), denoting the number of states per energy interval of the final state.

For beta decay processes, the matrix element \(V_{fi}\) is given by:

\begin{equation}
V_{fi} = g \int \psi_f^* \phi_e^* \phi_\nu^* O_x \psi_i \, dV
\label{matrix_element}
\end{equation}

where \(g\) is a constant determining the strength of the interaction. The final state consists of three components: \(\psi_f\), the wave function of the daughter nucleus (either in its ground state or an excited state, as constrained by the available Q value energy), and \(\phi_e\) and \(\phi_\nu\), which are the wave functions of the emitted electron and neutrino, respectively. The operator \(O_x\) denotes the interaction operator, while \(\psi_i\) is the wave function of the initial mother nucleus.

The wave functions for the electron and neutrino are approximated as those of free particles, normalized within volume \(V\):

\begin{equation}
\phi_e(\mathbf{r}) = \frac{1}{\sqrt{V}} e^{i \mathbf{p} \cdot \mathbf{r} / \hbar}, \quad \phi_\nu(\mathbf{r}) = \frac{1}{\sqrt{V}} e^{i \mathbf{q} \cdot \mathbf{r} / \hbar}
\label{free_particle}
\end{equation}

where \(\mathbf{p}\) and \(\mathbf{q}\) are the momenta of the emitted electron and neutrino, respectively. For electrons with energies around 1 MeV and \(r\) on a nuclear scale, these exponential terms can be expanded as:

\begin{equation}
e^{i \mathbf{p} \cdot \mathbf{r} / \hbar} = 1 + i \frac{\mathbf{p} \cdot \mathbf{r}}{\hbar} + \ldots
\end{equation}

with a similar form for the neutrino. The leading term of the expansion, \(1\), corresponds to \(r=0\), representing the scenario where the electron and neutrino are created at the origin and possess no orbital angular momentum. If the integration over the first term does not nullify its contribution, it becomes the primary contributor to the transition rate. If this term is zero, higher-order terms, indicative of non-zero orbital angular momentum, must be considered, resulting in a reduced transition rate.

In \(\beta\) decay, total angular momentum conservation establishes the selection rules. Both the \(\beta\) particle and neutrino are fermions with spin \(1/2\). When their orbital angular momentum is zero, any final angular momentum arises solely from their spins, which can be parallel (\(S=1\)) or antiparallel (\(S=0\)). The antiparallel alignment, termed \textit{Fermi decay}, results in no change in total angular momentum (\(\Delta I = 0\)) between the initial and final nuclear states. Alternatively, parallel alignment, known as \textit{Gamow-Teller decay}, allows for a quanta of angular momentum to be carried by spins, resulting in \(\Delta I = 0\) or \(1\), except when both \(I_i\) and \(I_f\) are zero.

The relationship between parity and angular momentum is \(\Delta \pi = (-1)^l\), where \(\pi\) denotes parity and \(l\) orbital angular momentum. The first expansion term presumes \(l=0\), implying \(\Delta \pi = 1\), which signifies no parity change between the initial and final nuclear states. This situation refers to \textit{allowed beta decay}, characterized by \(\Delta I = 0\) or \(1\) and \(\Delta \pi = \text{no}\).

When the first term's contribution nullifies, transitions involving non-zero orbital angular momentum become relevant. This phenomenon is known as \textit{forbidden decay}. Although the term 'forbidden' suggests impossibility, these decays are merely less probable and result in lower transition rates. When $\Delta \pi \neq \text{no}$, indicating a parity change, the condition for an allowed decay is unmet. To achieve $\Delta \pi \neq \text{no}$, the angular momentum must be $l = 1, 3, 5, \ldots$, with the lowest $l$ having the highest probability among these scenarios. Thus, $l = 1$ represents the most probable decay mode apart from the allowed decay process and is commonly referred to as \textit{first forbidden decay}. This occurs for $\Delta I = 0, 1, 2$ and $\Delta \pi = \text{yes}$.

The next probable scenario involves $l=2$, for which $\Delta \pi = \text{no}$, and $\Delta I$ could be $0, 1, 2, 3$. However, $\Delta I = 0, 1$ cases are often dominated by allowed decays, and their contribution from this term is minor. Consequently, significant contributions arise for $\Delta \pi = \text{no}$ and $\Delta I = 2, 3$, classified as \textit{second forbidden decays}. This process continues, leading to \textit{third forbidden decays}, and so forth. Each subsequent forbidden level has an approximate probability of $10^{-4}$ relative to the previous level.

To summarize, Table \ref{tab:selection_rules} provides the conditions associated with allowed, and first to third forbidden decays.

\begin{table}[h]
    \centering
    \begin{tabular}{llll}
        \toprule
        Decay Type & $\Delta I$ & Parity Change & Log $ft$ Range \\
        \midrule
        Superallowed & 0 & No change & $\sim$ 3 - 4 \\
        Allowed & 0, 1 & No change & 3.5 - 7.5 \\
        First Forbidden & 0, 1, 2 & Change & 6 - 9 \\
        Second Forbidden & 2, 3 & No change & 10 - 13 \\
        Third Forbidden & 3, 4 & Change & 14 - 20 \\
        \bottomrule
    \end{tabular}
    \caption{Summary of selection rules and associated log $ft$  (see section \ref{sec:log-ft}) values for different \(\beta\) decay types.}
    \label{tab:selection_rules}
\end{table}

For completeness, it should be noted that in some instances, the wave functions of the mother and daughter nuclear states are remarkably similar, except for a proton-neutron replacement. These scenarios, where $\Delta \pi = \text{no}$ and $\Delta I = 0$, are referred to as \textit{superallowed decays} and exhibit very high transition rates.

\subsection{\label{sec:log-ft}Comparative Half-lives}

Equation \ref{golden_rule} is revisited to examine the impact of the density of final states in beta decay processes. The number of states for an electron with momentum \( p \), between \( p \) and \( p + \Delta p \), is described by:

\begin{equation}
dn_e = \frac{V}{h^3} \cdot 4\pi p^2 \, dp
\end{equation}

assuming confinement within a volume \( V \). A similar expression is applied to the neutrino with momentum \( q \). Therefore, the combined number of states for the electron and neutrino is given by:

\begin{equation}
d^2 n = dn_e \, dn_\nu
\end{equation}

These expressions, along with \ref{matrix_element} and \ref{free_particle} are substituted into Equation \ref{golden_rule}, resulting in:

\begin{equation}
d\lambda = \frac{2\pi}{h} \cdot g^2 \lvert M_{fi} \vert^2 (4\pi)^2 \frac{p^2 \, dp \, q^2}{h^6} \cdot \frac{dq}{dE_f}
\end{equation}

where \( M_{fi} \) is represented by the integral:

\[
M_{fi} = \int \psi_f^* O_x \psi_i \, dv
\]

Integration over all possible \( p \) and \( q \) is performed to obtain the total rate of transition. Through momentum conservation, \( p \) is related to \( q \), allowing integration only over \( p \):

\begin{equation}
\lambda = \frac{g^2 \lvert M_{fi} \rvert^2}{2\pi^3 h^7 c^3} \int_0^{P_{\text{max}}} F(Z',p) \, p^2 (Q - T_e)^2 \, dp
\label{rate}
\end{equation}

Here, \( F \) represents the \textit{Fermi function}, which accounts for nuclear Coulomb field effects, since emitted electrons are not entirely free particles. The function \( F \) depends on \( Z' \), the atomic number of the daughter nucleus.

The term \( p^2 (Q - T_e)^2 \) is identified as the statistical factor derived from the finite available states for the electron and neutrino, with \( T_e \) representing the kinetic energy of the electron and \( Q \) the decay energy. The integral term in \ref{rate} focuses solely on \( Z' \) and the maximum electron energy \( E_0 \), defining the \textit{Fermi integral}:

\[
f(Z', E_0) = \frac{1}{m^5 c^7} \int_0^{P_{\text{max}}} F(Z',p) \, p^2 (E_0 - E_e)^2 \, dp
\]

Constants are included to ensure the integral remains dimensionless. By using the relation \(\lambda = \frac{0.693}{t_{1/2}}\), the following expression is derived:

\[
ft_{1/2} = 0.693 \cdot \frac{2 \pi^3 \hbar^7}{g^2 m_e^5 c^4 \lvert M_{fi} \rvert^2}
\]

This term is known as the \textit{comparative half-life} or \textit{ft value}, which enables the comparison of \(\beta\) decay probabilities across different nuclei with varying half-lives. The dependence of the ft value on \( M_{fi} \) indicates that changes result from alterations in the nuclear wave function. As ft values span an extensive range, they are often expressed as \(\log_{10} ft\) values.

Since ft is inversely proportional to \(\lvert M_{fi} \rvert^2\), representing the transition probability, higher ft values correspond to lower transition probabilities. Decays with the shortest comparative half-lives (\(\log ft \approx 3 - 4\)) are classified as the superallowed decays. Generally, typical \(\log ft\) values characterize different levels of forbiddenness, as detailed in Table \ref{tab:selection_rules}.

\subsection{\(\beta\) Delayed Emissions}

As previously discussed, \(\beta\) decay results in the population of excited states of the daughter nuclei. These excited states are often unstable and can decay by emitting \(\gamma\) rays as they transition to their ground state. Additionally, in certain cases, these states may also result in the emission of nucleons.

For \(\beta\) decay to lead to nucleon emission, the energy released during the decay, known as $Q_\beta$+, must exceed the separation energy of the nucleon from the daughter nucleus. This situation allows for the possibility of nucleon emission occurring alongside \(\gamma\) decay. The emission of a nucleon transitions the nucleus into a different nucleus, thus altering its identity.

This scenario is particularly relevant for our experiment involving the decay of $^{23}\text{Al}$. The $Q_\beta$+ energy of $^{23}\text{Al}$ is 11199.7 (3) keV ,greater than the proton separation energy of its daughter nucleus, $^{23}\text{Mg}$, which is $S_P$ = 7581.25 (14) keV \cite{nndc}. As a result, we anticipate observing both \(\beta\)-delayed \(\gamma\) emissions and proton emissions during the decay process in this experiment.

\section{\(\gamma\) Decay Theory}

\( \gamma \) decay involves the emission of \( \gamma \) rays, which are high-energy photons released as a nucleus transitions from a higher energy state to a lower one. Unlike \( \beta \) decay, \( \gamma \) decay does not change the number of protons or neutrons within the nucleus; rather, it occurs within the same isotope as the nucleus reaches its ground state configuration following other decay processes or nuclear reactions. \( \gamma \) rays possess a high penetration capability, requiring dense materials such as lead or several centimeters of concrete for adequate shielding.

The energy of the photon emitted by the  \( \gamma \) decay corresponds to the energy difference between the initial excited nuclear level and the final state of the nucleus post-decay, with an additional correction for nuclear recoil. This recoil energy can be determined using the principles of energy and momentum conservation, and be expressed as:

\begin{equation}
E_\gamma = \delta_E \left(1 - \frac{1}{2} \frac{\delta_E}{M_d c^2}\right),
\end{equation}

where \(E_\gamma\) denotes the energy of the emitted \(\gamma\), \(\delta_E\) is the actual energy difference between the initial and final excited states, \(M_d\) is the mass of the nucleus, and \(c\) is the speed of light. The equation shows that this correction becomes more significant with larger energy differences between the initial and final states. In our experiment, calculated correction values reached up to 1.5 keV for the higher \(\gamma\) energies, indicating that this correction is substantial and was therefore included in the calibration process and the final assessment of the excited levels. 

The relationship between the measured \(\gamma\) energy and the energy of the excited states allows for the exploration of excited energy levels within nuclei through the measurement of the energies of the emitted photons.

The emitted \(\gamma\) rays, as a form of electromagnetic radiation, can be classified according to the electromagnetic multipoles involved in their emission. Specifically, the multipoles include electric multipoles such as the dipole (E1) and quadrupole (E2), as well as magnetic multipoles such as the dipole (M1) and quadrupole (M2) \cite{blatt1979theoretical}. Each of these multipole fields is associated with distinct angular momentum values, quantified by the angular momentum \( l \) per emitted photon in units of \( \hbar \).

\subsection{Selection Rules}

The conservation of angular momentum during \(\gamma\) decay imposes selection rules that dictate which multipoles can participate in specific gamma transitions between two nuclear states. The possible angular momentum of the emitted photon is constrained by the following condition:

\begin{equation}
|I_f - I_i| \leq l_\gamma \leq I_f + I_i
\label{eq:angular_momentum}
\end{equation}

where \( I_f \) and \( I_i \) represent the angular momentum of the final and initial states, respectively, and \( l_\gamma \) denotes the angular momentum of the emitted photon. In cases where \( I_i = I_f \), only quadrupole or higher multipoles can participate in the transition. Specifically, transitions for \( I_i = I_f = 0 \) are entirely forbidden.

Additionally, parity conservation is an essential principle in \(\gamma\) emission processes. The parity associated with electric multipoles is related to their angular momentum by the equation:

\begin{equation}
\pi = (-1)^l
\label{eq:electric_parity}
\end{equation}

Conversely, for magnetic multipoles, parity is defined as:

\begin{equation}
\pi = (-1)^{l+1}
\label{eq:magnetic_parity}
\end{equation}

As a result, if there is a parity change between the initial and final states, only odd electric multipoles and even magnetic multipoles are permitted. Conversely, when there is no parity change, even electric multipoles and odd magnetic multipoles are allowed.

To summarize the selection rules governing gamma decay, the following equation encapsulates the conditions pertaining to angular momentum and parity:

\begin{equation}
    \begin{array}{c}
        |I_f - I_i| \leq l_\gamma \leq I_f + I_i \\
        \Delta \pi = \text{change} : \text{odd } E, \text{ even } M \\
        \Delta \pi = \text{no change} : \text{even } E, \text{ odd } M 
    \end{array}
\label{eq:selection_rules_summary}
\end{equation}

The ability of electric multipoles to radiate power surpasses that of their magnetic counterparts. Consequently, if allowed, the electric multipole contribution dominates over the magnetic contribution. Furthermore, the strength of radiation decreases for higher electromagnetic multipoles, which affects the rate of transitions based on the participating multipoles. The relationships governing these rate changes can be summarized as follows:

\begin{equation}
    \begin{array}{c}
        \frac{\lambda (E,l)}{\lambda (M,l)} \approx 10^2 \\[0.5em]
        \frac{\lambda (E,l+1)}{\lambda (E,l)} \approx 10^{-5} \\[0.5em]
        \frac{\lambda (M,l+1)}{\lambda (M,l)} \approx 10^{-5}
    \end{array}
\label{eq:transition_rates}
\end{equation}

%% file: 5.experimental_setup.tex
\chapter{Experimental Setup}
\label{chap:experimental_setup}

\section{Introduction}

The purpose of this chapter is to describe the experimental setup used to study the decay of \(^{23}\text{Al}\). This setup was implemented at the National Superconducting Cyclotron Laboratory (NSCL) at Michigan State University. The experiment aimed to investigate the decay properties of \(^{23}\text{Al}\), a proton-rich isotope, through its \(\beta\)-delayed proton emission and \(\gamma\)-ray spectroscopy.

The experimental configuration comprised the production of a radioactive beam, its subsequent transport to the detection systems, and the specific arrangements for detecting decay products with high precision and efficiency.

The following sections detail the production of the \(^{23}\text{Al}\) beam, the comprehensive detection system, and the operational procedures employed to maximize data quality.

\section{Production of the \(^{23}\text{Al}\) Beam}

The \(^{23}\text{Al}\) beam utilized in this experiment was produced at the NSCL via the process of projectile fragmentation. A primary beam of \(^{36}\text{Ar}\) ions, accelerated to an energy of 150 MeV/u using the Coupled Cyclotron Facility \cite{Marti2001}, was directed onto a \(^{9}\text{Be}\) target with a thickness of 1363 mg/cm\(^2\). This interaction resulted in the fragmentation of the primary beam, yielding a variety of isotopic products, including \(^{23}\text{Al}\).

To isolate the \(^{23}\text{Al}\) from other fragments, the beam was passed through the A1900 magnetic fragment separator \cite{Stolz2005}. The A1900 is a versatile instrument that utilizes strong magnetic fields to separate ions based on their mass-to-charge ratio. After initial separation, a Radio Frequency Fragment Separator (RFFS) \cite{Bazin2009} was employed to enhance the purity of \(^{23}\text{Al}\), achieving a purity level of approximately 69\%.

\begin{sidewaysfigure}
    \centering
    \includegraphics[width=\textheight]{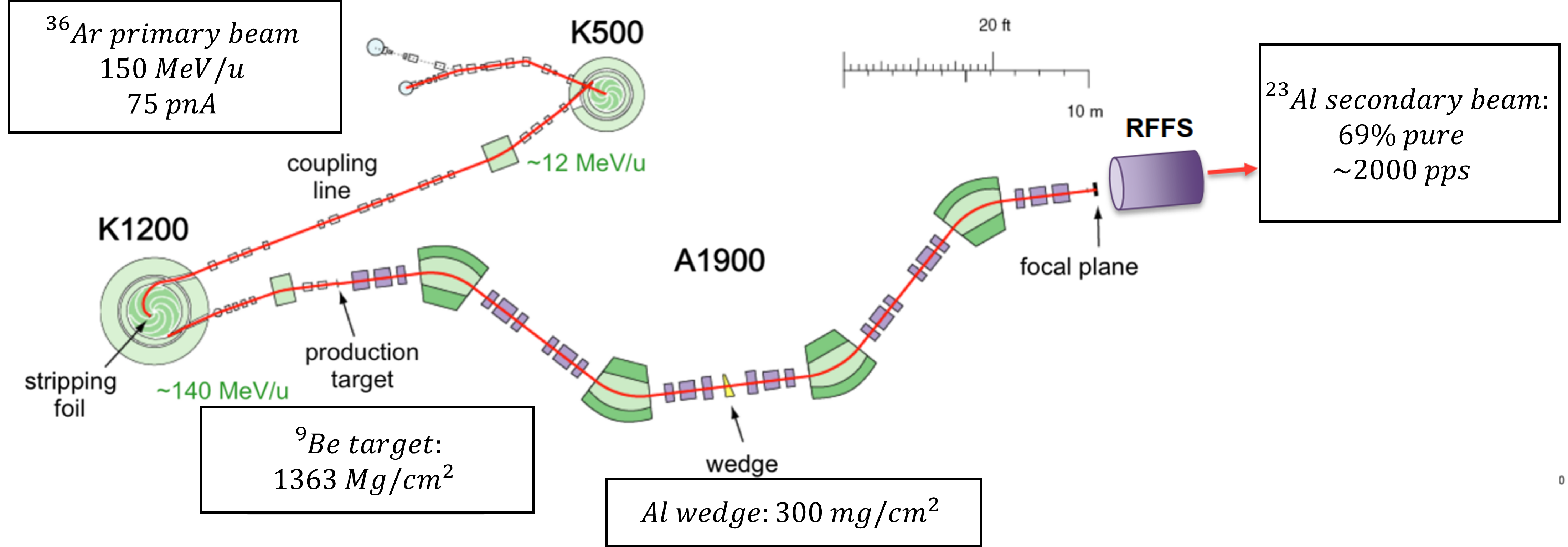}
    \caption{Schematic of the NSCL beam line and the \(^{23}\text{Al}\) creation process.}
    \label{fig:NSCL_beam_line}
\end{sidewaysfigure}

Following separation, the \(^{23}\text{Al}\) beam was transported downstream to the detector setup. The beam was analyzed for contaminants using a silicon detector placed after the RFFS, which is located 6 meters upstream of the detector. The primary contaminants, in decreasing order of intensity, were identified as \(^{21}\text{Na}\), \(^{22}\text{Mg}\), and \(^{16}\text{N}\), utilizing the standard \(\Delta E\) time-of-flight method. The rate of \(^{23}\text{Al}\) delivery to the experimental apparatus was approximately 2000 ions per second, providing a sufficient flux for detailed decay studies \cite{Friedman2020}.

\section{GADGET Detection System}

The Gaseous Detector with Germanium Tagging (GADGET) detection system is meticulously designed to facilitate the simultaneous measurement of \(\gamma\) rays and low energy \(\beta\)-delayed protons \cite{Friedman2019}. This system integrates advanced technologies to achieve high efficiency in capturing and analyzing decay events. At the core of the GADGET system, the Segmented Germanium Array (SeGA) and the proton detector are crucial components, working in unison to provide complementary measurements.

The SeGA surrounds the proton detector to achieve high-efficiency gamma detection. The integration of these components allows for the efficient capture of coincident decay products, enabling detailed insight into the decay chains of the isotope.

\begin{figure}[h]
    \centering
    \includegraphics[width=0.8\textwidth]{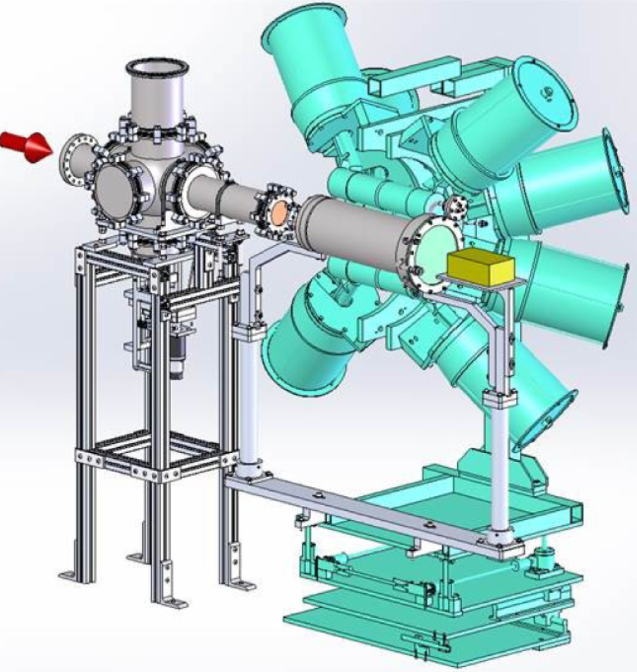}
    \caption{GADGET assembly showing the beam path (red arrow) entering the degrader cross and depositing \(^{23}\text{Al}\) into the proton detector. The proton detector is surrounded by the SeGA (colored Cyan) for coincidence \(\gamma\) detection. Only half of SeGA is shown for clarity. Adopted from Friedman et al. \cite{Friedman2019}.}
    \label{fig:GADGET_assembly}
\end{figure}

In addition to SeGA and the proton detector, the detection system includes a beam-pipe cross, containing a degrader for fine-tuning of the beam energy and a silicon detector for particle ID.

\subsection{Proton Detector}

\begin{figure}[h]
    \centering
    \includegraphics[width=0.8\textwidth]{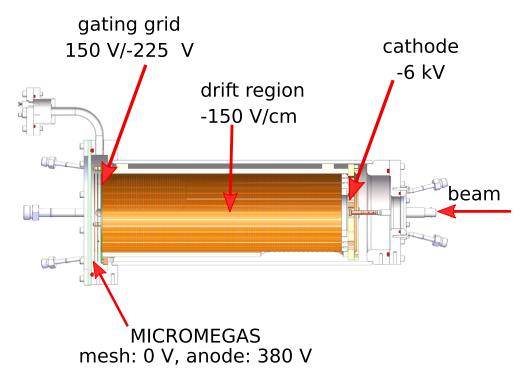}
    \caption{Cross-section of the proton detector. The effective detection area measures 40 cm in length and 10 cm in diameter. Adopted from Friedman et al. \cite{Friedman2019}.}
    \label{fig:GADGET_proton_detector}
\end{figure}

The proton detector, a central component of the GADGET system, is designed to capture \(\beta\)-delayed protons emitted during decay processes with high precision. The ionized gas within the proton detector serves as both the stopping medium for the beam particles and the detection medium for measuring proton energies, with its active length sufficiently covering the dispersion of the stopping range in the gas. The proton detector measures 40 cm in length and 10 cm in diameter, as illustrated in Fig. \ref{fig:GADGET_proton_detector}.

The detector utilizes a P10 gas mixture at a pressure of 780 Torr, where incoming protons ionize the gas, resulting in the creation of electron-ion pairs. These liberated electrons are directed by an applied electric field of 150 V/cm towards the readout plane, where they are amplified using the MICROMEGAS (MICRO MEsh GAseous Structure) array \cite{Giomataris2006}.

\begin{figure}[h]
    \centering
    \includegraphics[width=0.8\textwidth]{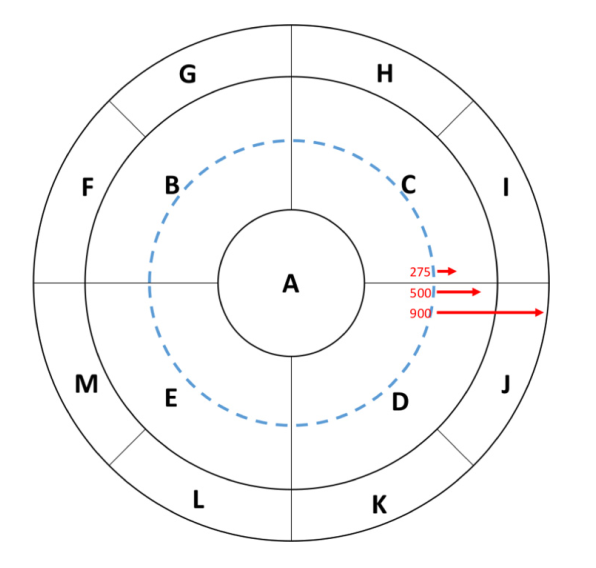}
    \caption{Geometry of the detection pads on the anode plane of the proton detector. The radii are 1.4, 4, and 5 cm. A dashed line marks the cylindrical radius of 2.6 cm limiting the beam dispersion. Proton ranges of 275, 500, and 900 KeV are marked by red lines. Schematic figure adopted from Friedman et al. \cite{Friedman2019}.}
    \label{fig:GADGET_pads_ranges}
\end{figure}

The MICROMEGAS plane is divided into 13 pads marked A-M, facilitating a veto mechanism for high-energy protons that may exit the active medium without depositing all their energy. This division allows comprehensive analysis of multiple pad readings for enhanced detection efficiency or focuses on a single pad for reducing beta noise and improving energy resolution, since the pads combined spectrum is limited by the resolution of the pad with the poorest performance. Notably, during the experiment, pads F, G, L, and M were not implemented, which reduced the veto efficiency to 50\%. However, this limitation is not significant for the \(^{23}\text{Al}\) case as only a few protons exceed 1 MeV \cite{Kirsebom2011}, ensuring that any resulting background, due to partial energy measurement of high-energy protons, is minimal. An aperture at the cathode limits the dispersion of the beam to a radius of 2.6 cm around the mutual axis of the beam and the detector, and the pads are organized with the radii of 1.4, 4, and 5 cm (see Figure~\ref{fig:GADGET_pads_ranges}).

Constructed with a cylindrical geometry and positioned concentrically with the beam axis, the detector's design optimizes its functionality for effective beam moderation and precise proton detection. The detection rate is limited to an order of \(10^{4}\) events per second under nominal working conditions. This limitation arises from the electron drift time, which depends on factors such as the gas mixture, pressure, and electric field.

The design of the proton detector emphasizes transparency to \(\gamma\) rays, employing materials specifically chosen to minimize absorption. This ensures that while protons are accurately measured, \(\gamma\) rays can pass through efficiently to be detected by the surrounding SeGA. Additionally, the detector operates under specific vacuum conditions to minimize gas contamination. A continuous gas flow is maintained to reduce gas contaminates from outgassing, which would hinder electron transport towards the readout plane.

\subsection{SeGA}
The Segmented Germanium Array (SeGA) is employed for the detection of \(\gamma\) rays emitted during the decay processes \cite{Mueller2001}. Each detector within SeGA is designed to deliver high-resolution \(\gamma\)-ray spectroscopy, measuring 0.3\% at 1332 KeV, ensuring the capture of intricate details about the decay of \(^{23}\text{Al}\).

The Germanium detectors are composed of high-purity Germanium crystals. When a \(\gamma\)-ray photon enters a germanium crystal, it interacts with the material typically through the photoelectric effect, Compton scattering, or pair production. These interactions result in the ionization, forming electron-hole pairs. An applied electric field across the germanium causes these charges to move, resulting in an electric current pulse that is measured to determine energy accurately.

SeGA comprises 16 coaxial, high-purity Germanium detectors. The Germanium detectors are arranged in two rings of eight detectors each: one set positioned upstream and the other downstream relative to the center of the active volume of the proton detector (see Figure~\ref{fig:GADGET_assembly}). This configuration allows SeGA to provide comprehensive spatial coverage of emitted \(\gamma\) rays, maximizing detection efficiency and facilitating coincident measurement capabilities as part of the GADGET system.

\section{Beam Operation and Measurement Procedure}

The beam functioned in a pulsed mode to efficiently manage the accumulation and measurement cycles essential for capturing decay events.

During the beam-on phase, the beam was directed into the detector and allowed to accumulate \(^{23}\text{Al}\) over a 0.5-second interval. Following this, the beam was turned off, initiating a 0.5-second beam-off period during which measurements of \(\beta\)-delayed protons were conducted. To protect the Micromegas from the strong direct beam signal, a gating grid was activated during beam-on periods. In contrast to the proton detector, \(\gamma\) rays were measured by the SeGA during both beam phases, allowing for continuous data collection throughout the entire experiment. This cyclical approach enabled the effective measurement of the relatively weak signal of the emitted protons without interference from the strong signal of the direct beam.

The overall experiment consisted of 11 measurement runs, each lasting approximately 1 hour to ensure comprehensive data acquisition. Additionally, several beam-off runs were conducted to collect background data.

%% file: 6.analysis_and_results.tex
\chapter{Data Analysis}
\label{chap:analysis_and_results}

This chapter presents the data analysis and results derived from the experimental measurements. The analysis was conducted using ROOT, a data analysis framework, allowing for efficient handling and visualization of the complex datasets.

\section{\(\gamma\) Peak Fitting Method}

\begin{figure}[H]
    \centering
    \includegraphics[width=0.8\textwidth]{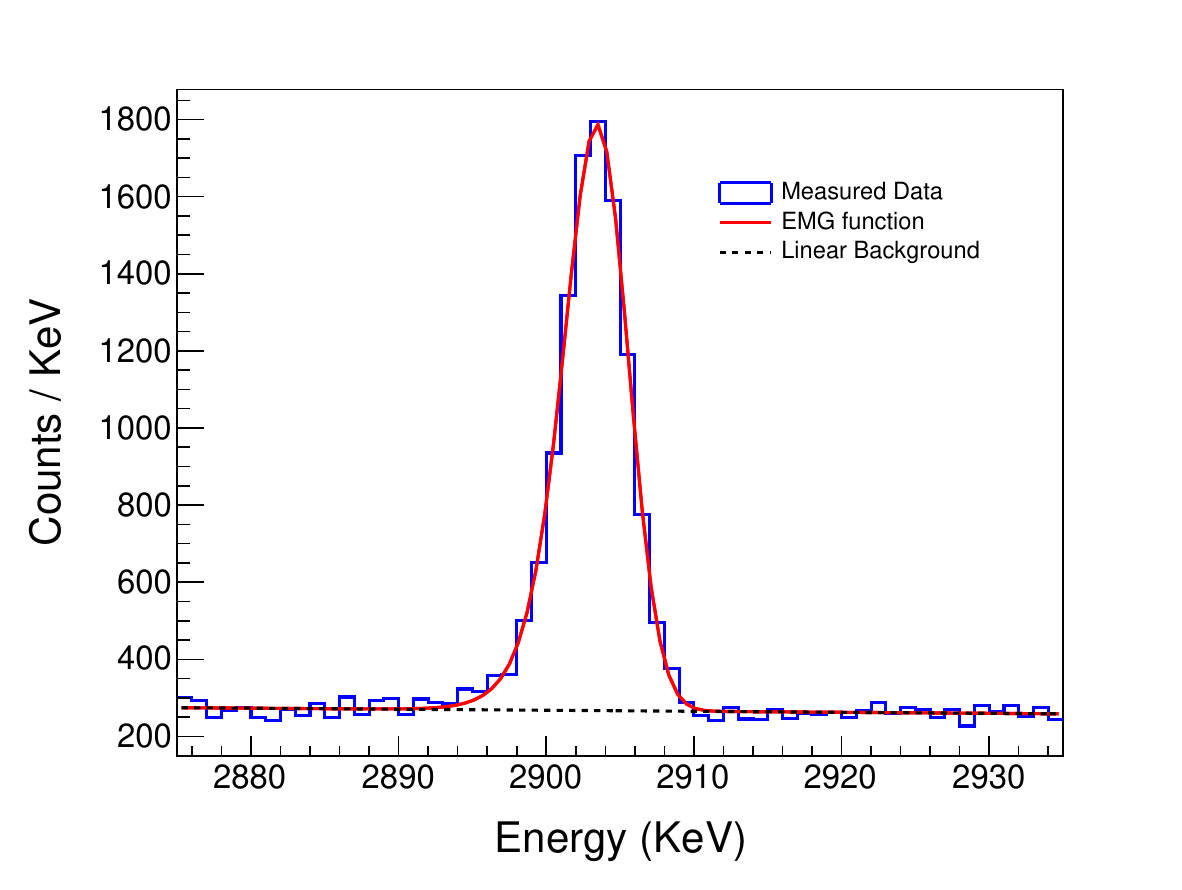}
    \caption{Example of a gamma peak fitted using the EMG function combined with a linear background model. The fitted curve demonstrates the characteristic wider tail on the lower energy side of the peak and the linear behavior of the surrounding background.}
    \label{fig:f1-peak_fitting_method}
\end{figure}

To analyze the \(\gamma\) peaks, an Exponentially Modified Gaussian (EMG) function  was employed to fit the observed data. The EMG function appropriately models the generally Gaussian nature of the gamma peak distributions while accounting for the asymmetries introduced by the detector's response function. Specifically, the broader tail observed on the lower energy side of the peaks is captured by this function. This asymmetry, which is illustrated in Figure \ref{fig:f1-peak_fitting_method}, arises primarily due to some ionized electrons escaping the detection volume, especially those originating near the detector's edges. As a result, there is a skew towards the left side of the peak's mean, leading to a slight shift of the perceived mean toward lower energies.

The EMG function account for these factors, offering a practical approximation for our objectives: to accurately determine the peak counts and mean energy. While not a flawless representation of the detector's complete response, the EMG function comprehensively meets the analysis requirements in this context. The EMG function is mathematically represented as follows:

\begin{align}
f(E; N, \mu, \sigma, \tau) = &\; \frac{N}{2\tau} 
\exp\left[\frac{1}{2} \left( \frac{\sigma}{\tau} \right)^2 + \frac{E - \mu}{\tau}\right] \notag \\
&\times \mathrm{erfc}\left[\frac{1}{\sqrt{2}}\left(\frac{\sigma}{\tau} + \frac{E - \mu}{\sigma}\right)\right]
\end{align}

where $\tau$ is the exponential decay constant, $\mu$ and $\sigma$ are the mean and the standard deviation of the Gaussian component, respectively, $E$ is the energy and $N$ is the total area under the curve. Initially, all four parameters (\(\tau\), \(\sigma\), \(\mu\), and \(N\)) were determined through fitting. However, an energy dependency for \(\tau\) and \(\sigma\) will be established later in the analysis (see section \ref{sec:energy_dependency}). Consequently, all final \(\gamma\) peak fits will be re-conducted with only two parameters being freely fitted, while \(\tau\) and \(\sigma\) will be fixed according to their obtained energy dependence.

Additionally, to account for the local background, a linear function was incorporated into the fitting procedure. The background generally exhibits linear behavior across most gamma peaks, with special adjustments made in cases where deviations were noted, as will be detailed later. All fits were performed using the minimum \(\chi\)-square method, except in cases where statistical limitations necessitated the use of maximum likelihood estimation.

\section{SeGA Detectors Performance Overview}

In this section, we provide a brief overview of the performance of the SeGA \(\gamma\) detectors utilized in the analysis. As outlined in the Experimental Setup chapter, the setup included a total of 16 SeGA detectors. However, not all detectors were employed for this analysis due to several issues.

One detector was excluded because it was non-functional. Another detector was not used because it was only capable of measuring \(\gamma\) energies up to approximately 5000 keV, whereas the others extended beyond 8000 keV. Since our analysis requires comparing \(\gamma\) count ratios between peaks, consistency across detectors is crucial, leading to the exclusion of this limited-range detector.

Additionally, four detectors exhibited poor resolution compared to their counterparts. Inclusion of these lower-resolution detectors would compromise the accuracy of the measured peak mean, an undesirable outcome for this analysis. Consequently, only 10 of the SeGA detectors were employed in the final data analysis.

\section{Channel - Energy Calibration of the SeGA Detectors}

To facilitate the creation of a cumulative \(\gamma\)-ray energy spectrum, gain-matching of the SeGA detectors for each measurement run was performed. Five strong and well-characterized \(\gamma\) peaks were selected for this purpose: three known peaks from $^{23}\text{Mg}$ at 451 keV, 1601 keV, and 7803 keV \cite{nndc}, along with two prominent background peaks at 1461 keV from $^{40}\text{K}$  \cite{Chen2017} and 2615 keV from $^{208}\text{Tl}$ \cite{Martin2007}.

\begin{figure}[H]
    \centering
    \includegraphics[width=0.8\textwidth]{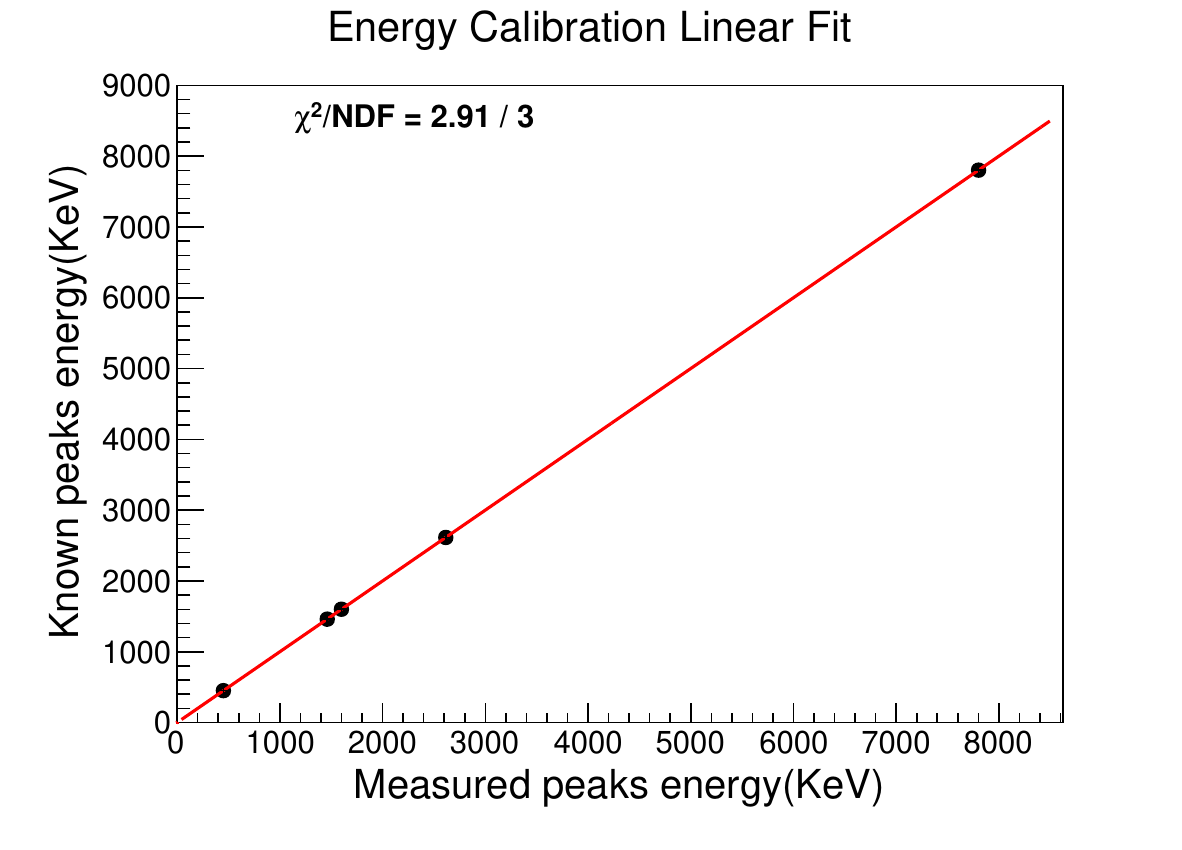}
    \caption{Example of a linear fit between known energy values and measured peak positions used for calibrating SeGA detectors. Although uncertainties were accounted for, the error bars are not visible in the figure due to their relatively small size.}
    \label{fig:f4_channel_energy_calibration_linear_curve_example}
\end{figure}

For each of the 11 measurement runs and for each of the 10 selected detectors, these five peaks were fitted. Figure \ref{fig:f4_channel_energy_calibration_linear_curve_example} presents an example of the linear relationship established between the known energy values and the corresponding measured peak positions, which allowed the derivation of the energy calibration parameters for each detector and run.

Subsequently, the data from each run and detector were individually calibrated using their respective calibration parameters. This calibration enabled the amalgamation of all individually calibrated measurements into a single, unified cumulative \(\gamma\)-ray spectrum. 

\begin{figure}
    \hspace*{-3.7cm} 
    \includegraphics[width=1.5\textwidth]{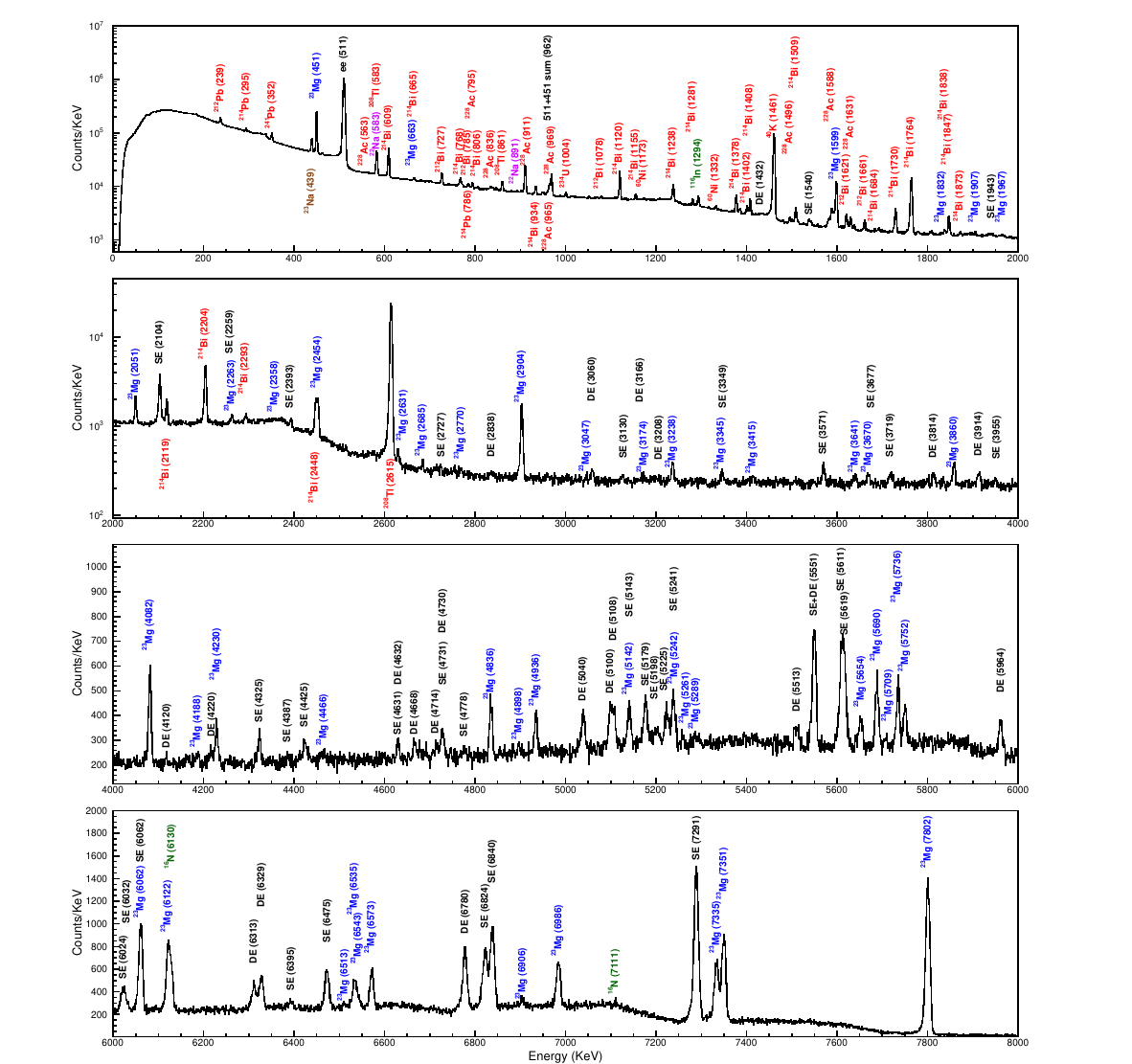}
    \caption{Unified \(\gamma\)-ray spectrum up to 8 MeV obtained from the calibrated data across all measurement runs and detectors. Identified \(\gamma\) rays are labeled as follows: \({}^{23}\text{Mg}\) transitions are represented in blue, \({}^{22}\text{Na}\) transitions in purple, and one \({}^{23}\text{Na}\) transition in brown. Background peaks are marked in red, while single escape (SE), double escape (DE), electron-positron annihilation, and one sum peak are labeled in black. Beam impurities are indicated in green.}
    \label{fig:f5_calibrated_ungated_spectrum}
\end{figure}

Finally, the unified gamma-ray spectrum, resulting from the calibrated and combined data of all runs and detectors, is shown in Figure \ref{fig:f5_calibrated_ungated_spectrum}.

\section{\(\gamma\) Peak Analysis}

The analysis of the \(\gamma\) spectrum focuses on identifying peaks associated with the \({}^{23}\text{Al}\) beam to construct a complete decay scheme. This process involves eliminating extraneous peaks from background radiation, escape peaks, and beam contamination, as well as evaluating decay chains from \(\gamma\) emissions following a prior \(\gamma\) decay or proton emission. This chapter describes the methodologies and tools used to examine \(\gamma\) peaks.

\subsection{\(\beta\) Gated Spectrum}

A \(\beta\)-\(\gamma\) coincidence spectrum was constructed to effectively isolate peaks generated by the \(^{23}\text{Al}\) beam. As \(^{23}\text{Al}\) decays via \(\beta\) emission to \(^{23}\text{Mg}\), subsequent decays can occur either through \(\gamma\) emission to lower excited states or to the ground state of \(^{23}\text{Mg}\), proton emission to \(^{23}\text{Na}\), or another \(\beta\) decay to \(^{23}\text{Na}\) potentially accompanied by \(\gamma\) emission (see Fig. \ref{fig:isotopes}). While examining this spectrum, it is crucial to note that beam contamination peaks may also appear. To rule out these contaminant peaks, we identified known strong \(\gamma\) peaks associated with contaminants and marked them as such.

\begin{figure}
    \centering
    \includegraphics[width=\textwidth]{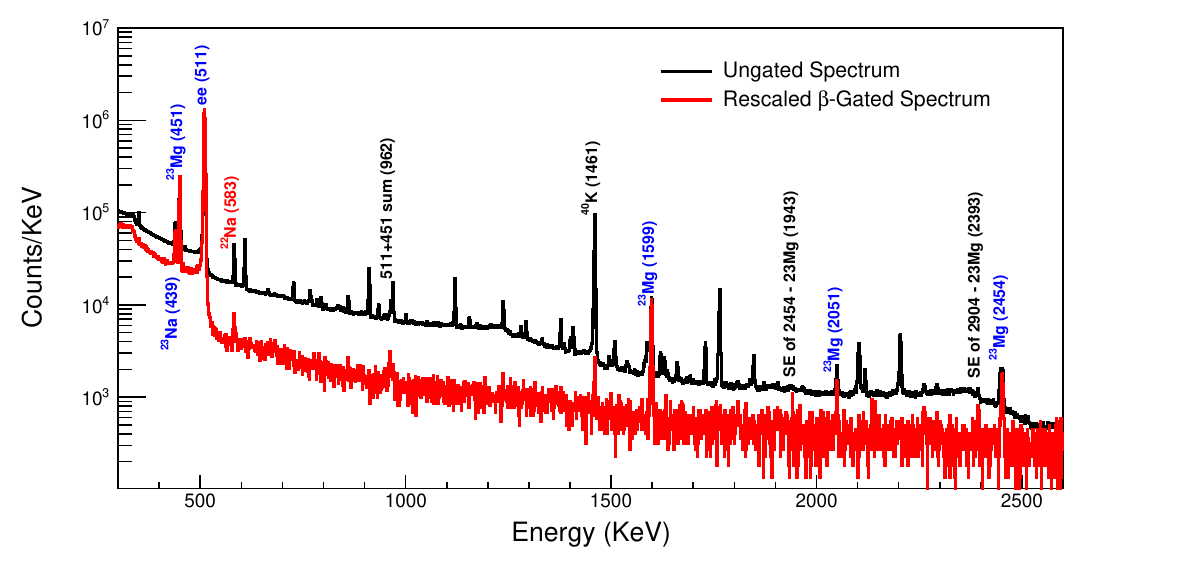}
    \caption{Illustration of a portion of the \(\beta\)-gated \(\gamma\)-ray spectrum. Identified \({}^{23}\text{Al}\) \(\beta\)-delayed \(\gamma\) rays are labeled. However, not all \(\beta\)-delayed \(\gamma\) rays appear prominently in the gated spectrum; the relatively weak \(\beta\)-delayed \(\gamma\) rays are often obscured by lower efficiency caused by the \(\beta\) gating, which is compounded by the already low statistics. Room background peaks present in the ungated spectrum (black spectrum, unlabeled peaks) are notably absent in the gated spectrum. The strong 1461 keV \({}^{40}\text{K}\) room background peak appears in the gated spectrum due to high random coincidence, but it is significantly more attenuated compared to the true \(\beta\)-delayed \(\gamma\) rays.}
    \label{fig:ungated_vs_gate}
\end{figure}

The \(\beta\)-gated spectrum was constructed through time gating, leveraging both energy and time measurements of events. Only \(\gamma\) events that occurred immediately following a \(\beta\) event detected by the proton detector were included in the spectrum. A coincidence time window of 8 \(\mu\)s was established for this process, corresponding to the full detector length drift time. The resulting spectrum is presented in Fig. \ref{fig:ungated_vs_gate}.

\(\beta\) gating substantially reduces detection efficiency due to below-threshold energy deposition of \(\beta\) particles in the Proton Detector, along with the gating-grid cycle's efficiency. However, any known strong background peaks that may appear in the \(\beta\)-gated spectrum from random coincidences are much more attenuated compared to the \(\beta\)-delayed \(\gamma\) rays. This attenuation ultimately results in a cleaner \(\gamma\) spectrum. It should be noted that the \(\beta\)-gated spectrum was used solely for peak identification, while the actual analysis was conducted with the ungated spectrum due to its superior statistics.

\subsection{Background Spectrum}

Another tool used to rule out non-beam originated peaks was the background spectrum. This spectrum was constructed using data collected during several measurement sessions when the beam was inactive. Naturally, peaks originating from the beam should not appear in the background spectrum, which should only contain room background peaks. The background measurements were calibrated and gain-matched using the same channel-energy calibration method described earlier for the beam measurements. The calibration of the background spectrum utilized five specific peaks: 583.187 ± 0.02 keV from \(^{208}\text{Pb}\), 609.317 ± 0.05 keV from \(^{214}\text{Po}\), 1460.820 ± 0.005 keV from \(^{40}\text{K}\), 1120.294 ± 0.06 keV from \(^{214}\text{Po}\), and 2614.522 ± 0.01 keV from \(^{208}\text{Pb}\).

Since the background measurements were conducted over a shorter timeframe than the beam-on measurements, the background spectrum counts were normalized to match the duration of the beam-on measurements for clarity of examination.

\begin{figure}[H]
    \centering
    \includegraphics[width=\textwidth]{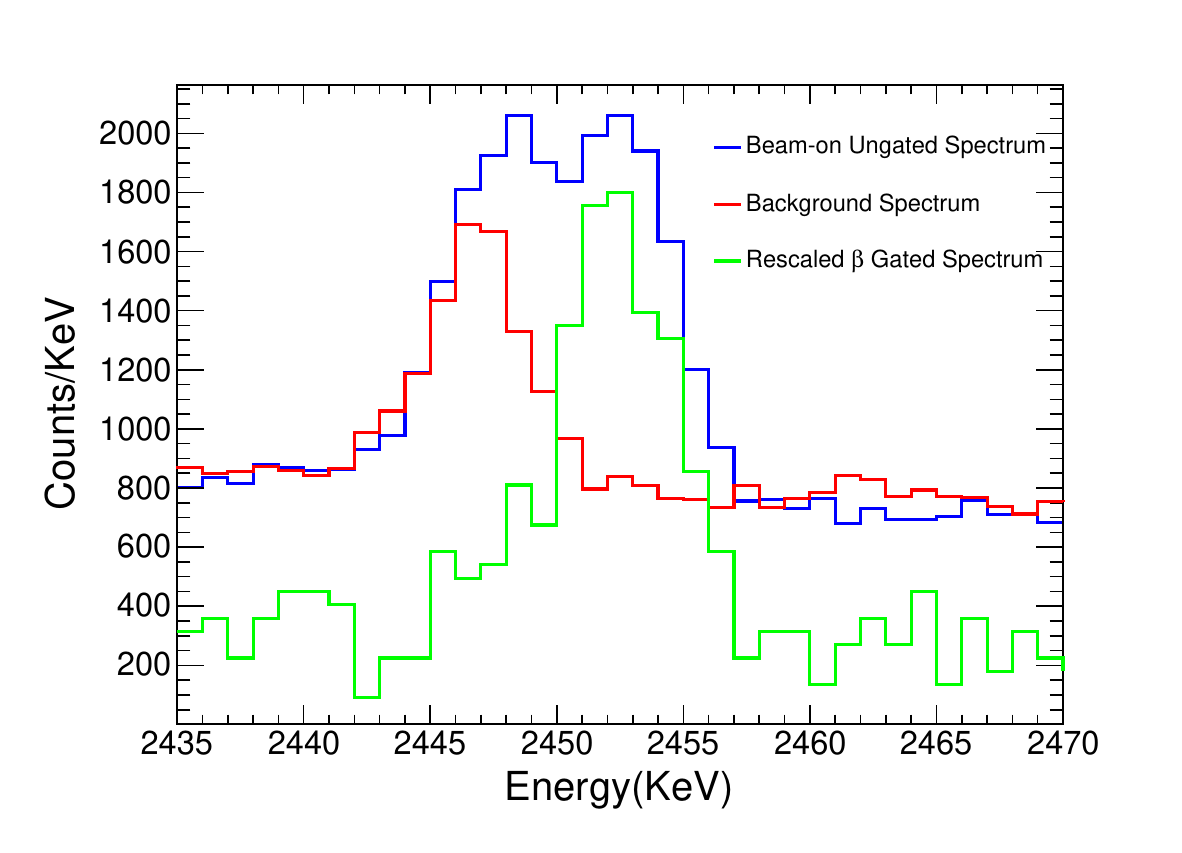}
    \caption{Peak analysis around 2450 keV. The ungated \( \gamma \) spectrum reveals peak splitting, while the \( \beta \)-gated spectrum confirms the presence of the right peak and the background spectrum highlights the left peak.}
    \label{fig:split_peak_example}
\end{figure}

An illustrative example involves a peak observed at approximately 2450 keV in the \(\gamma\) spectrum, as seen in Figure \ref{fig:split_peak_example}. The splitting of the peak's center suggests the presence of two closely positioned peaks. Examination of the \(\beta\)-gated spectrum reveals that only the right peak appears there, while the background (BG) spectrum shows that only the left peak is present. This observation allows us to conclude that there are indeed two peaks, one originating from the beam and the other from the background.

\subsection{\(\gamma-\gamma\) Coincidences}

The \(\gamma-\gamma\) coincidence technique is a valuable tool for examining decay chains and gaining a deeper understanding of the excited states in \(^{23}\text{Mg}\). This approach helps distinguish between decays that transition directly to the ground state and those that proceed through lower excited states. A two-dimensional histogram for \(\gamma-\gamma\) coincidences was constructed using a time gate to facilitate this analysis. For each measured \(\gamma\) event, any subsequent \(\gamma\) event occurring within a \(\pm 200\) ns time window was added to the histogram.

This time window is significantly longer than the characteristic lifetimes of the excited states, yet considerably shorter than the expected time separation between two distinct decay chain events. The discrepancy in processing times across different SeGA detectors, typically around 200 ns, is the primary reason for requiring a relatively broad time window. While the processing time of the \(\gamma\) detector causes the signal to register later than the actual event, the time window chosen ensures ample coverage to account for this variability. A window of \(\pm 200\) ns offered the optimal real coincidence to random coincidence ratio and was therefore selected for the \(\gamma-\gamma\) coincidence analysis.

\begin{figure}
    \centering
    \includegraphics[width=\textwidth]{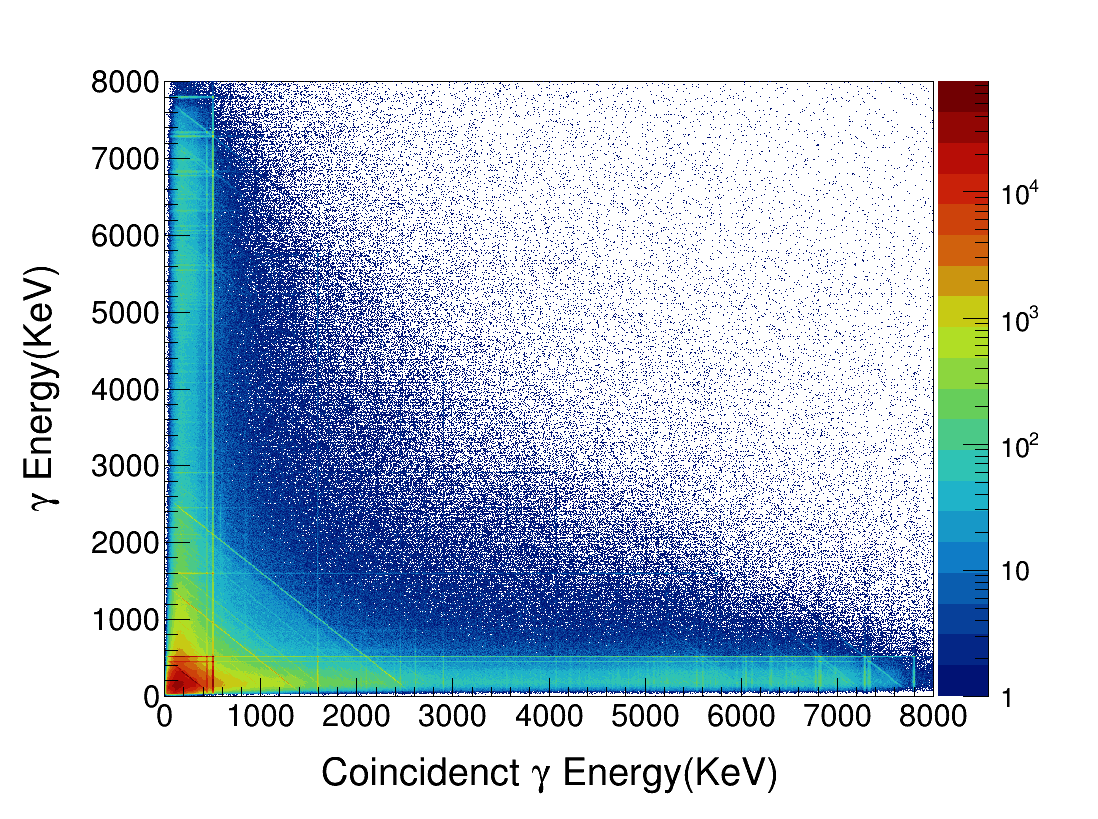}
    \caption{Two-dimensional \(\gamma-\gamma\) coincidence spectrum with a time window of \(\pm 200\) ns. The histogram illustrates various coincidence events, including diagonal lines resulting from energy sharing between detectors, and straight lines due to intense peaks with high random coincidences.}
    \label{fig:gg_coincidence_histogram}
\end{figure}

Figure \ref{fig:gg_coincidence_histogram} illustrates the two-dimensional \(\gamma-\gamma\) coincidence histogram constructed with this 200 ns time window. Within the figure, coincidences can be examined and analyzed by projecting onto one of the axes and subtracting background noise. A notable feature of the histogram is the presence of diagonal coincidence lines. These diagonals arise from a unique phenomenon: a single \(\gamma\) ray enters one detector, deposits part of its energy, escapes, and then deposits the remaining energy in another detector. Consequently, these diagonals represent instances where the sum of energies on the x and y axes equals the actual energy of the \(\gamma\) ray. Such diagonals are not indicative of true \(\gamma-\gamma\) coincidences but rather stem from the experimental setup and its inherent limitations.

In addition to diagonals, the histogram also shows straight lines across the plot. These lines result from intense peaks that generate numerous random coincidences. To assess whether a coincidence at a certain energy is genuine, a two-dimensional noise reduction technique will be employed to better distinguish real coincidences from background noise.

Table \ref{tab:gamma_coincidences} displays the \( \gamma \) rays that were measured in coincidence. Some expected \( \gamma \)-\( \gamma \) coincidences were not detected, likely due to low statistics. Specifically, these undetected coincidences involve small peaks, which resulted in insufficient data for conclusive identification of \( \gamma \)-\( \gamma \) coincidences.

\begin{table}[H]
    \caption{\( \gamma \) rays that were detected in coincidence with each other during the experiment.}

    \label{tab:gamma_coincidences}
    \centering
    \begin{multicols}{2} 
    \begin{tabular}{@{}c *{5}{>{\centering\arraybackslash}c}@{}}
        \toprule
        \textbf{E\textsubscript{\(\gamma\)} (KeV)} & \textbf{451} & \textbf{1600} & \textbf{2455} & \textbf{} & \textbf{2906} \\
        \midrule
        451  &   & \checkmark &   &  & \\
        663  &   & \checkmark &   &  & \\
        1599 & \checkmark &  &   &  & \\
        1832 &   &   &   &  & \\
        1907 &   &   &   &  & \\
        1967 &   &   &   &  & \\
        2051 &   &   &   &  & \\
        2263 & \checkmark &  &   &  & \\
        2358 &   &   &   &  & \\
        2454 & \checkmark &  &   &  & \\
        2631 &   &   &   &  & \\
        2685 &   &   &   &  & \\
        2770 &   &   &   &  & \\
        2904 &   &   &   &  & \\
        3047 &   &   &   &  & \\
        3174 &   &   &   &  & \\
        3238 &   & \checkmark &  &  & \\
        3415 &   &   &   &  & \\
        3641 &   &   &   &  & \\
        3670 &   &   &   &  & \\
        3860 &   &   &   &  & \\
        4082 & \checkmark &  & \checkmark &  & \checkmark \\
        4188 &   &   &   &  & \\
        4230 & \checkmark &  &   &  & \\
        \end{tabular}
    \begin{tabular}{@{}c *{5}{>{\centering\arraybackslash}c}@{}}
        \toprule
        \textbf{E\textsubscript{\(\gamma\)} (KeV)} & \textbf{451} & \textbf{1600} & \textbf{2455} & \textbf{} & \textbf{2906} \\
        \midrule
        4466 &   &   &   &  & \\
        4836 & \checkmark &  &   &  & \\
        4898 &   &   &   &  & \\
        4936 &   & \checkmark &  &  & \\
        5142 &   &   &   &  & \\
        5242 & \checkmark &   &   &  & \\
        5261 &   &   &   &  & \\
        5289 &   &   &   &  & \\
        5654 &   &   &   &  & \\
        5690 &   &   &   &  & \\
        5709 &   &   &   &  & \\
        5736 & \checkmark & \checkmark &  &  & \\
        5752 & \checkmark & \checkmark &  &  & \\
        6062 & \checkmark &  &   &  & \\
        6122 & \checkmark &  &   &  & \\
        6513 &   &   &   &  & \\
        6535 & \checkmark &   &   &  & \\
        6543 &   &   &   &  & \\
        6573 &   &   &   &  & \\
        6906 &   &   &   &  & \\
        6986 &   &   &   &  & \\
        7335 & \checkmark &   &   &  & \\
        7351 & \checkmark &   &   &  & \\
        7802 &   &   &   &  & \\
    \end{tabular}
    \end{multicols} 
\end{table}

\subsubsection{Example: \(\gamma-\gamma\) Coincidence Contribution to Decay Chain Analysis}

The \(\gamma-\gamma\) coincidence analysis provides valuable insights into the decay chains, as demonstrated in the following example. A \(\gamma\) peak was observed at 4083 keV. This peak did not appear in the background spectrum and is not a known \(\gamma\) ray transition associated with \(^{23}\text{Mg}\), its \(\beta\) decay daughter \(^{23}\text{Na}\), or any known beam contaminant. Initially, one might hypothesize that this peak indicates a previously undiscovered excited state in \(^{23}\text{Mg}\).

\begin{figure}[H]
    \centering
    \includegraphics[width=\textwidth]{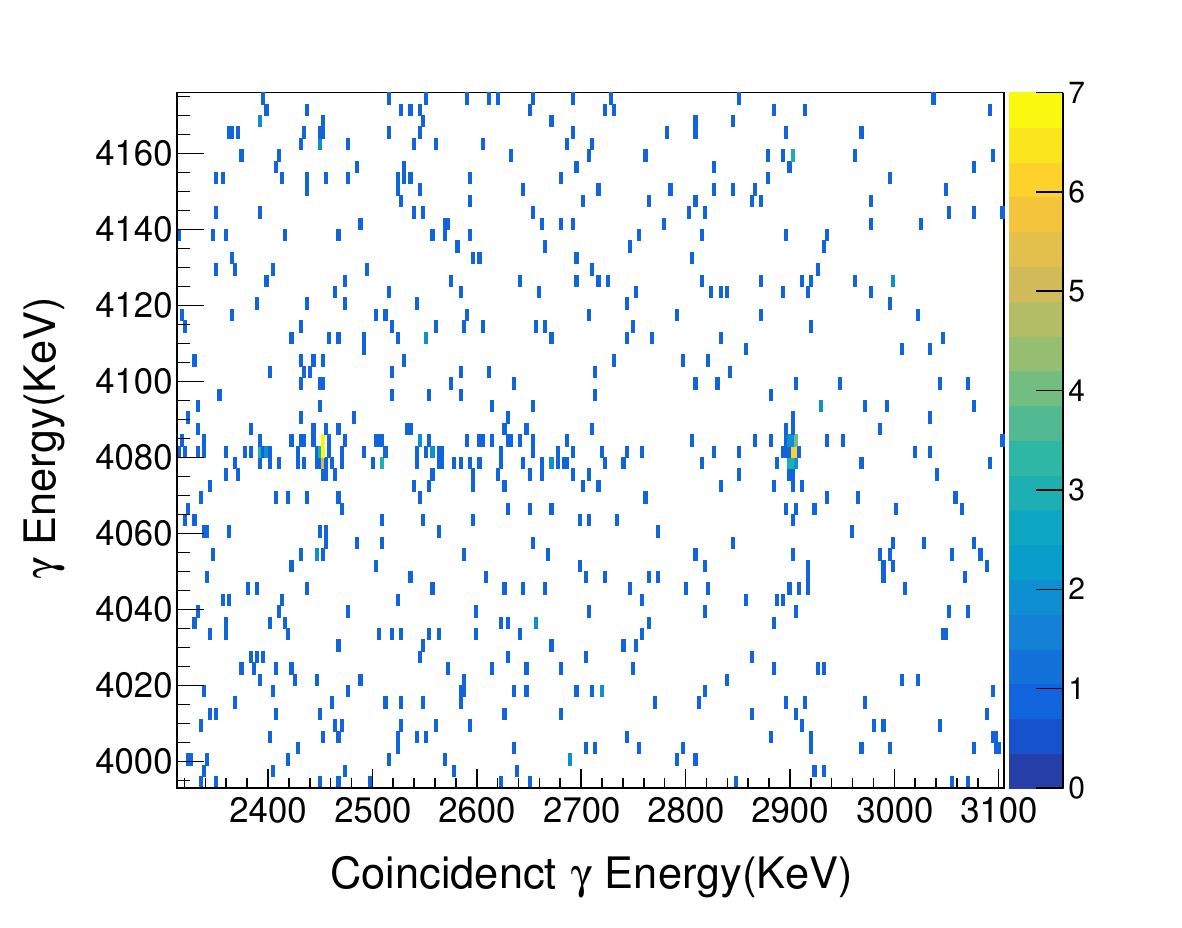}
    \caption{Zoom-in on the 4082 keV \(\gamma-\gamma\) coincidences with 2904 keV and 2454 keV, illustrating its origin from the 6984 ± 5 keV excited state. A two-dimensional background subtraction was used to verify those coincidences.}
    \label{fig:example_gg_coincidence}
\end{figure}

However, as illustrated in \ref{fig:example_gg_coincidence}, a careful examination of the \(\gamma-\gamma\) coincidences reveals that the 4082 keV \(\gamma\) is coincident with both the 2454 keV and 2904 keV \(\gamma\) rays, both known transitions of \(^{23}\text{Mg}\). The 2904 keV and 2454 keV \(\gamma\) rays originate from the 2904 keV excited level, with transitions from 2904 keV to the ground state and 2904 keV to the first excited state at 451 keV, respectively. Consequently, we can infer that the 4082 keV \(\gamma\) likely originates from a higher energy state decaying into the 2904 keV state. The known excited state at 6984 ± 5 keV matches this scenario. Although \(\gamma\) rays from this state were not observed in previous experiments, it is now evident that the 4082 keV \(\gamma\) originates from this level. Thus, no evidence supports the existence of an excited level at 4082 keV itself.

\subsection{Proton-Gamma Coincidence}

As previously discussed, the ability to measure proton-\(\gamma\) coincidences is a significant advantage of the GADGET detector system. By utilizing both the proton detector and SeGA, a proton-\(\gamma\) coincidence spectrum was constructed. The process was analogous to the \(\gamma-\gamma\) coincidence method but used a time window of 8 \(\mu\)s.

This coincidence measurement primarily aimed to investigate potential \(\gamma\) peaks from \(^{22}\text{Na}\) excited states. If these peaks occur, they are expected to follow a proton emission from \(^{23}\text{Mg}\) and thus be detected in coincidence with the emitted proton. In this experiment, two \(\gamma\) rays associated with \(^{22}\text{Na}\) were identified, which will be further discussed later in this chapter.

\begin{figure}[H]
    \centering
    \includegraphics[width=\textwidth]{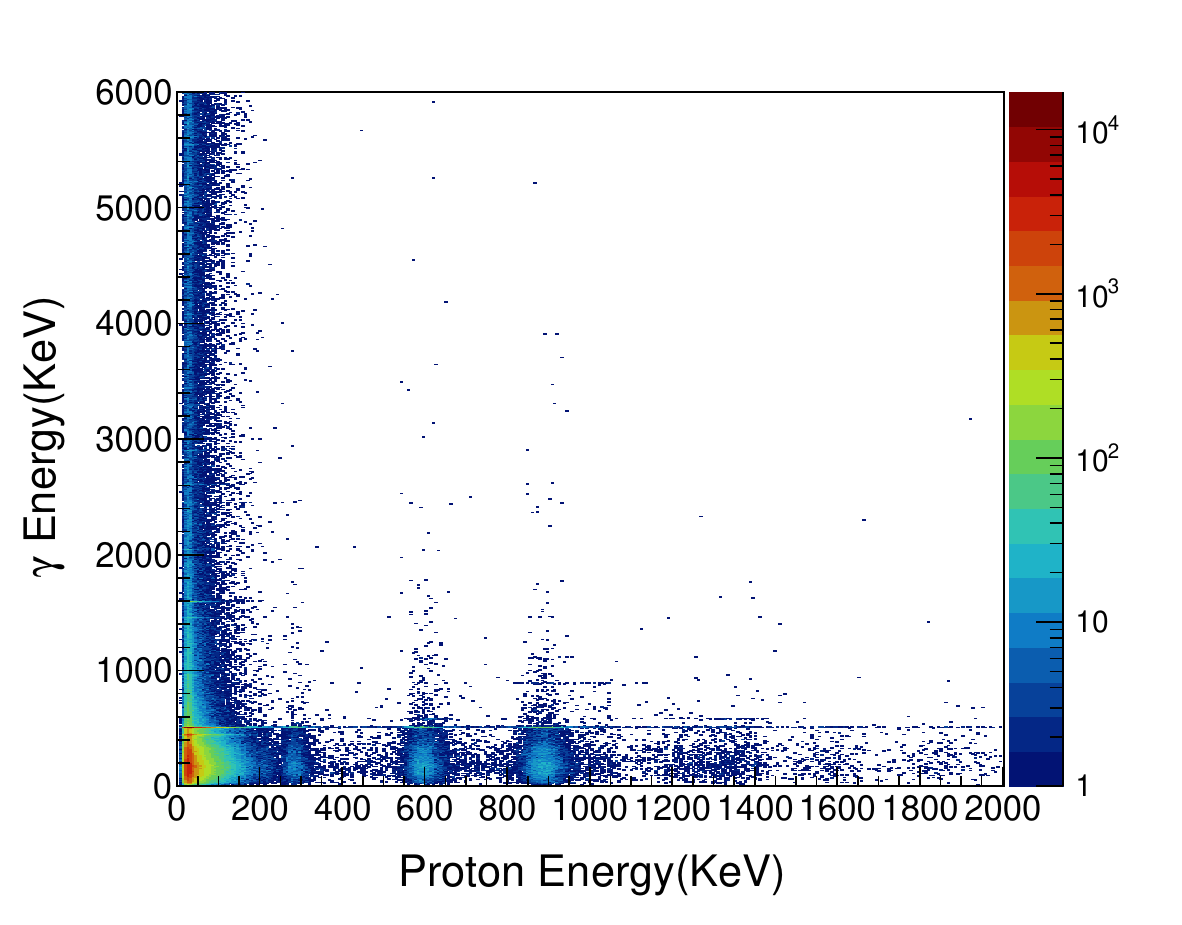}
    \caption{Measured two-dimensional proton-\(\gamma\) coincidence spectrum, using a time window of 8 \(\mu\)s.}
    \label{fig:pg_coincidence_histogram}
\end{figure}

Figure \ref{fig:pg_coincidence_histogram} presents the two-dimensional proton-\(\gamma\) coincidence spectrum with a time window of 8 \(\mu\)s, illustrating the experimental results.

\section{\(\gamma\) Detection Efficiency}

\subsection{\label{sec:sim_eff}Simulated Relative Efficiency Curve}

The relative efficiency of the SeGA detectors was determined using a GEANT4-based simulation \cite{Agostinelli2003}. This simulation considered both the \(\gamma\)-ray photopeak detection efficiency of the detectors and the spatial geometrical configuration, integrating the geometry of the experimental setup and the spatial distribution of the beam.

\begin{figure}
    \centering
    \includegraphics[width=0.8\textwidth]{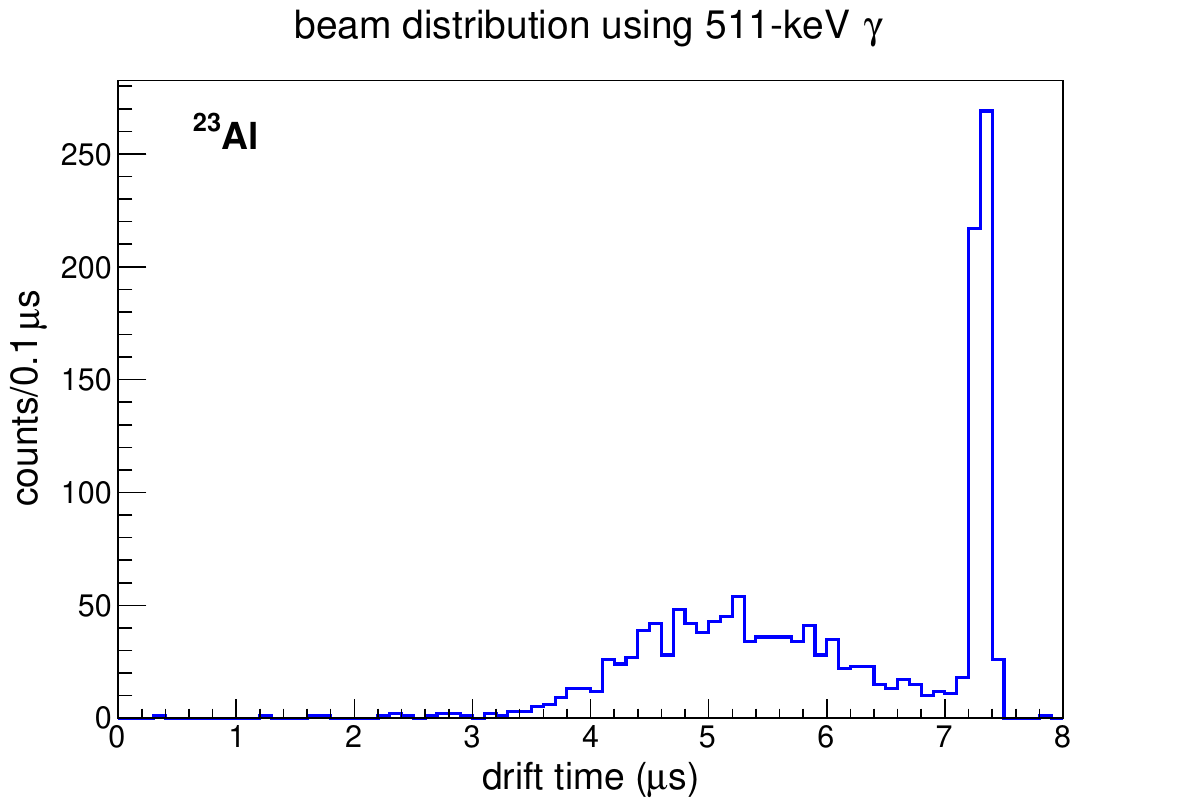}
    \caption{Longitudinal distribution of the beam in the detector, showing two peaks with one at the detector's edge. This illustrates the electron reabsorption phenomenon by ionized nuclei.}
    \label{fig:beam_longitudinal_distribution}
\end{figure}

The efficiency depends on each Germanium detector's inherent properties and their spatial arrangement. The 3D spatial distribution of the beam was characterized through two main analyses. The time distribution of ionized electron arrival at the proton detector's readout plane was measured, producing a one-dimensional distribution of the beam along the detector's length. This distribution, depicted in Figure \ref{fig:beam_longitudinal_distribution}, reveals two peaks, one notably at the detector's edge. This phenomenon occurs as ionized nuclei initially lack electrons and subsequently absorb them from the gas. Nuclei capturing electrons quickly do not drift to the readout plane, whereas those with delayed capture contribute to the second peak due to drift caused by the electric field. This occurrence provides insights into the electron reabsorption timescale, though a detailed analysis extends beyond the scope of this work.

\begin{figure}
    \centering
    \includegraphics[width=0.7\textwidth]{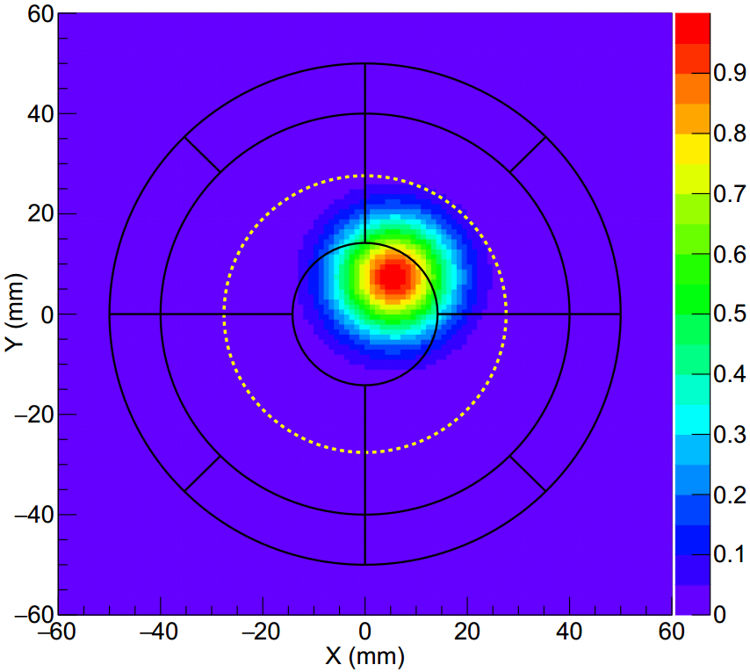}
    \caption{Two-dimensional spatial distribution of the beam based on MICROMEGAS pad events distribution.}
    \label{fig:beam_2D_distribution}
\end{figure}

By analyzing event distribution on the MICROMEGAS pads, the beam's positioning in the remaining two dimensions was inferred, as shown in Figure \ref{fig:beam_2D_distribution}.

These spatial distributions served as inputs to the GEANT4 simulation, which utilized a Monte Carlo approach to generate calculated efficiency values at 4 different energies of \(^{23}\text{Mg}\) \(\gamma\) rays. To these calculated efficiency data points, a function of the following form to obtain the efficiency curve:

\begin{align}
\text{Efficiency} = \exp\left(P_0 + P_1 \log(E) + P_2 \log^2(E)\right)
\end{align}

Based on previous experience with SeGA, this curve accurately reflects efficiency behavior across different energies; however, there is a systematic scaling uncertainty on the order of 10\% associated with it, necessitating scaling to an absolute efficiency value at a specific energy point (see section \ref{sec:abs-SeGA_eff}).

\begin{figure}[H]
    \centering
    \includegraphics[width=0.8\textwidth]{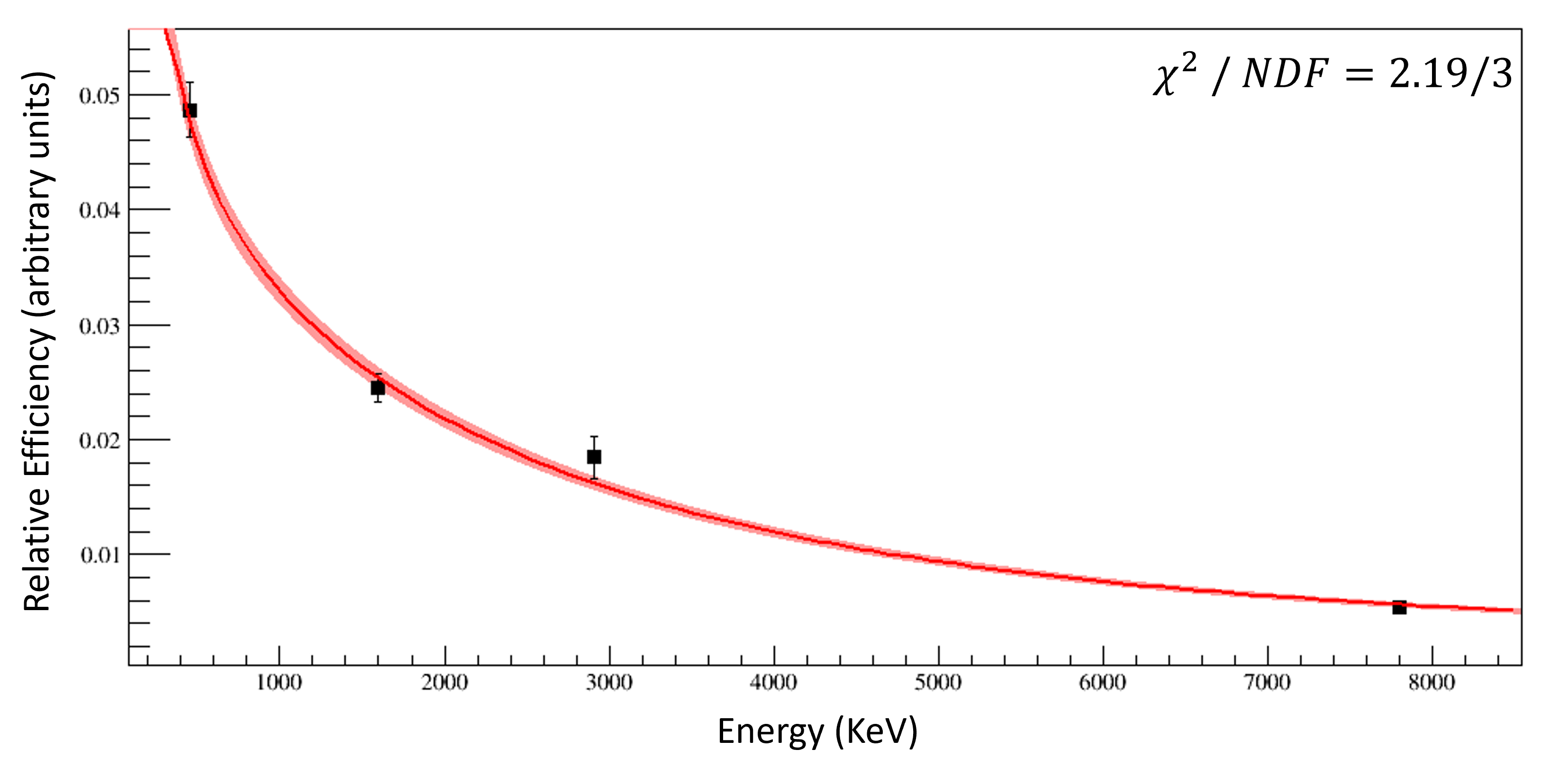}
    \caption{Simulated relative efficiency curve for the SeGA detectors, based on four \(^{23}\text{Mg}\) \(\gamma\) photopeaks.}
    \label{fig:simulated_efficiency_curve}
\end{figure}

Figure \ref{fig:simulated_efficiency_curve} illustrates the simulation results, displaying the relative efficiency curve derived from four \(^{23}\text{Mg}\) \(\gamma\) photopeaks.

A flat 2\% uncertainty was applied to the relative efficiencies at all energies to account for the \( \gamma \gamma \) summing effect \cite{Sun2021}. In addition, a conservative uncertainty envelope of 3\% is established for \( \gamma \)-ray energies below 1.4 MeV, while a 5\% uncertainty envelope is used for \( \gamma \)-ray energies above 1.4 MeV. The uncertainties associated with the relative efficiencies were propagated through the calculations for each \( \gamma \)-ray intensity.

\subsection{\label{sec:abs-SeGA_eff}Absolute Efficiency at 451KeV}

To determine the absolute efficiency of the SeGA detectors at a specific energy, we need a reference point to scale the relative efficiency curve, enabling calculation of absolute efficiency across all energies. For this purpose, we utilized certain known decay chains involving the 451 keV first excited state of \(^{23}\text{Mg}\). In these decay chains, every \(\gamma\) transition leads to another \(\gamma\) emission at 451 keV. For instance, the transition from the 2051 keV excited state to the 451 keV state emits a 1600 keV \(\gamma\), ensuring that each detected 1601 keV \(\gamma\) is followed by a 451 keV one.

We identified decay chains with such forced \(\gamma\)-\(\gamma\) coincidences and ensured that each transition involved a 3/2 angular momentum or lower, indicating isotropic emission. This is crucial to avoid spatial correlation effects that could influence efficiency analysis due to the detector array's spatial configuration.

\begin{figure}[H]
    \centering
    \includegraphics[width=\textwidth]{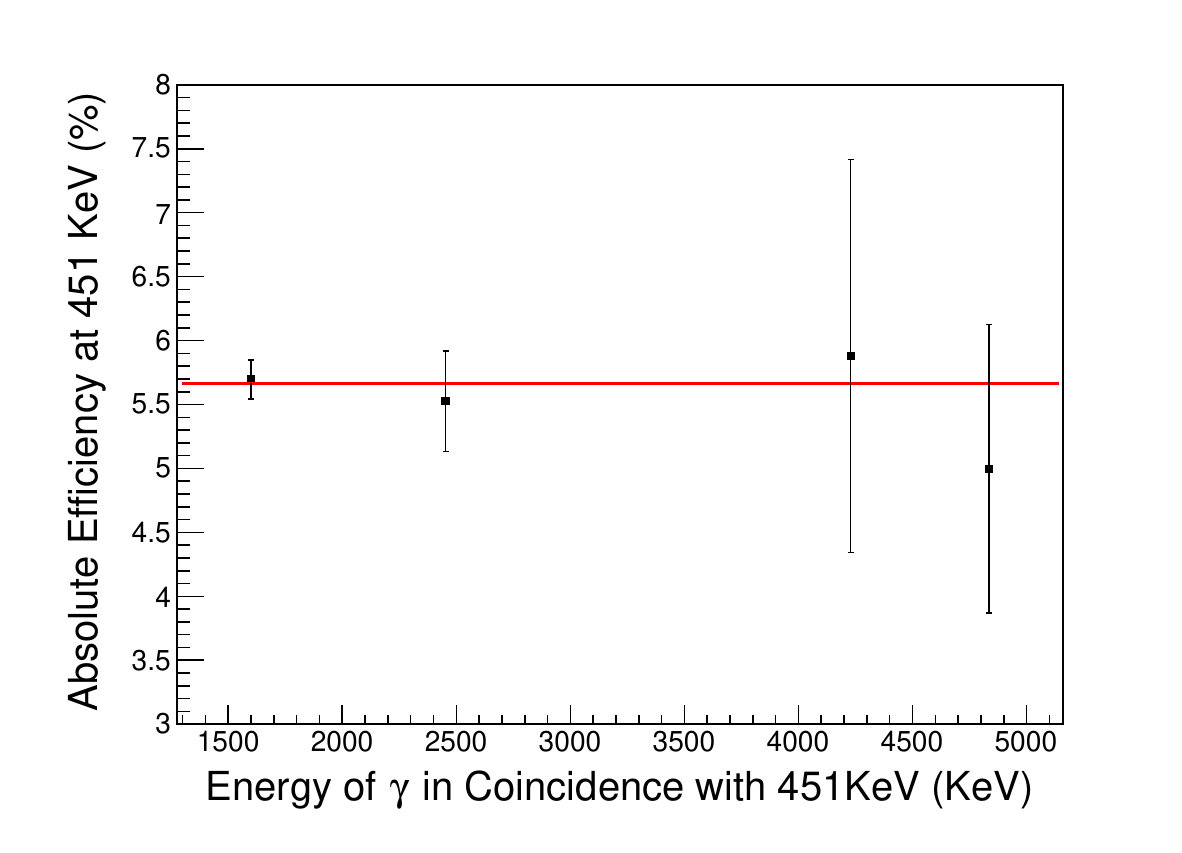}
    \caption{Ratios between detected \(\gamma\)-\(\gamma\) counts and ungated gamma peak counts for several peaks known to decay through the 451 KeV excited state. The fitted ratios determines the absolute efficiency at 451 KeV.}
    \label{fig:absolute_efficiency}
\end{figure}

By evaluating the ratio between counts of the original \(\gamma\) peak, such as the 1601 keV peak, and the \(\gamma\)-\(\gamma\) coincidence counts for 1600 KeV with 451 keV, we deduced the absolute efficiency at 451 keV. This process was applied to four known \(\gamma\)-\(\gamma\) coincidences passing through the 451 keV state: 1600, 2454, 4231, and 4837 keV. A fit was conducted to the ratios obtained for these four peaks, depicted in Figure \ref{fig:absolute_efficiency}, confirming their agreement and establishing an efficiency of \(5.66 \pm 0.14\%\).

Notably, in our setup, one of ten detectors serves as a trigger for the first \(\gamma\), whereas the remaining nine were used for efficiency measurement. If a subsequent \(\gamma\) was detected by the triggering detector, it would register the total energy, preventing coincidence detection and leading to an underestimation of efficiency. Therefore, we adjust the fitted number by a factor of \( \frac{10}{9} \), resulting in an absolute \(\gamma\) detection efficiency of \(6.30 \pm 0.16\%\).

\begin{figure}[h]
    \centering
    \includegraphics[width=\textwidth]{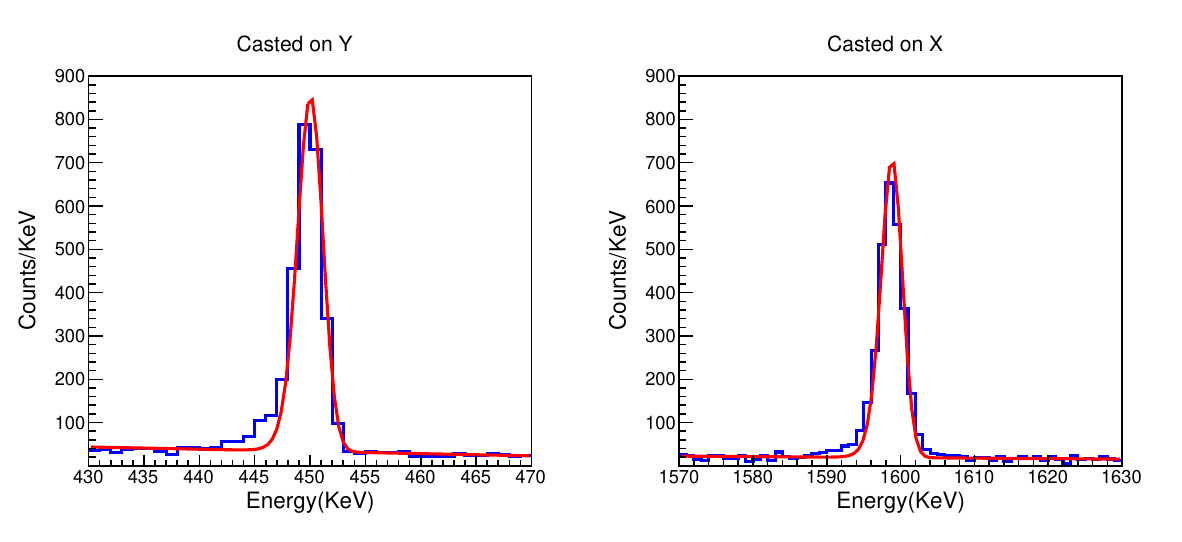}
    \caption{The left graph displays the fit of the 451 keV \(\gamma\) peak after projection of the 2D \( \gamma-\gamma \) spectrum onto the y-axis, while the right side shows the fit for the 1600 keV peak after projection of the 2D \( \gamma-\gamma \) spectrum onto the x-axis. The linear noise obtained from the 1600 keV fit was subtracted from the counts of the 451 keV peak to determine the real \( \gamma-\gamma \) count.}

    \label{fig:noise_reduction}
\end{figure}

During \(\gamma\)-\(\gamma\) coincidence peak fitting, we applied a two-dimensional noise reduction technique. The 2D histogram was projected onto both axes, subtracting the linear background of the primary \(\gamma\) (e.g., 1600) from the counts fitted for the 451 keV peak from the other axis. This process is illustrated in Figure \ref{fig:noise_reduction}.

\section{\label{sec:energy_dependency}Energy Dependency Analysis}

To enhance the accuracy of the EMG fit used in peak analysis, a reduction in the degrees of freedom associated with the fit parameters was aimed for. Specifically, an energy dependency was established for the parameters \(\tau\) and \(\sigma\), which describe the peak width and exponential decay, respectively. By understanding these dependencies, greater precision in determining the mean and area of the peaks, which are of primary physical interest, can be achieved.

After establishing these dependencies, all EMG fits to the \(\gamma\) peaks were reconducted, using the mean energies of the peaks to effectively eliminate \(\tau\) and \(\sigma\) as fit parameters.

For this analysis, several strong and well-known peaks were carefully selected. The process of determining the energy dependencies and the results obtained are discussed in the following subsections.

\subsection{\(\sigma\) Energy Dependency}

To describe the dependency of \(\sigma\) on energy, the following relation was employed:

\begin{equation}
\sigma(E) = \sqrt{a + bE + cE^2}
\end{equation}

where \(E\) is the mean energy of the measured peak. This function encapsulates various contributions to the resolution. The parameter \(a\) represents contributions from electronic noise and other constant factors that affect the resolution even at zero energy. These contributions are primarily due to the electronic components of the detection system. Meanwhile, parameter \(b\) accounts for the intrinsic resolution of the detector. The number of ions created within the detector is linearly dependent on the energy of the \(\gamma\) ray, following a Poisson distribution. As a result, the uncertainty in the number of ions, proportional to \(\sqrt{E}\), gives rise to an energy-dependent contribution to the resolution. Lastly, parameter \(c\) captures non-linearities in detector response, such as those arising from charge collection inefficiencies. This quadratic term becomes increasingly significant at higher energies, where non-linear effects more heavily impact detector accuracy.

\begin{figure}[H]
    \centering
    \includegraphics[width=\textwidth]{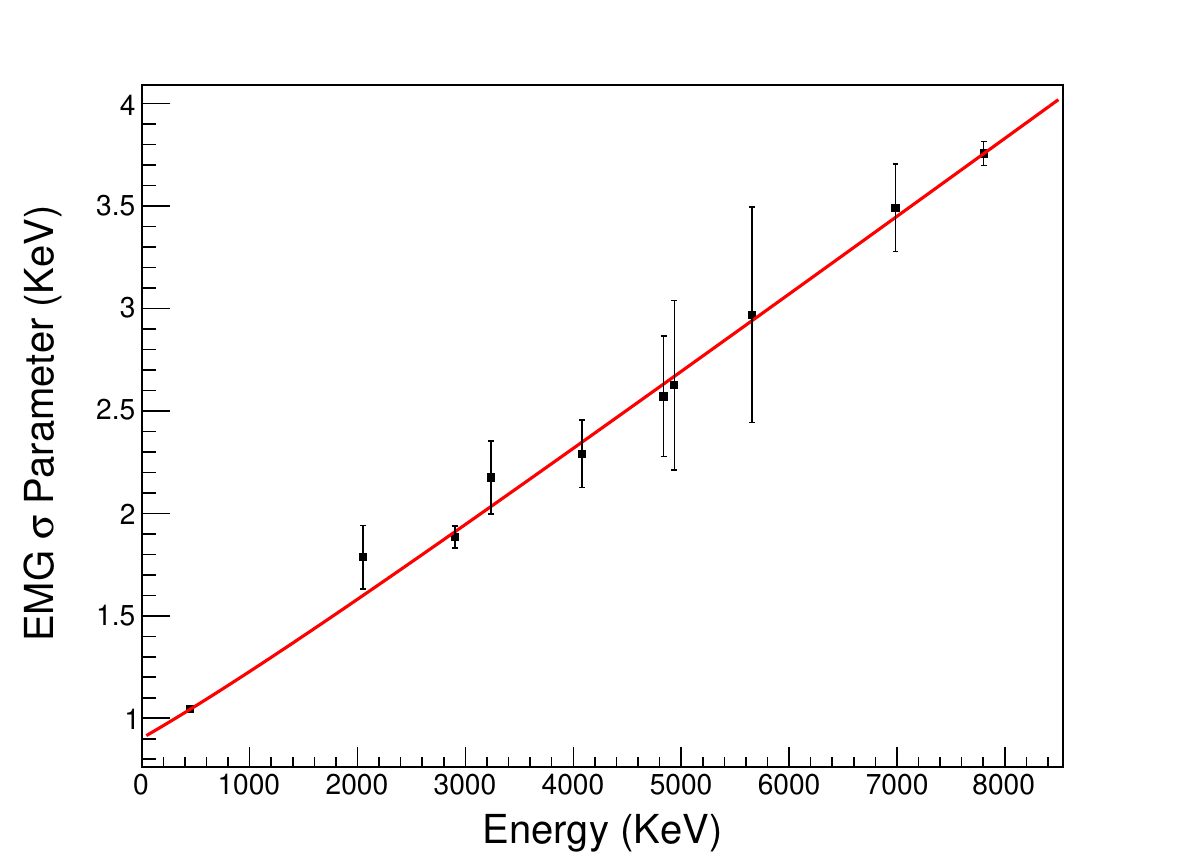}
    \caption{Energy depndency of the EMG \(\sigma\) parameter.}
    \label{fig:sigma_energy_fit}
\end{figure}

In Figure \ref{fig:sigma_energy_fit}, the fit of \(\sigma(E)\) to the experimentally measured \(\sigma\) values at various energies is illustrated. It is demonstrated that this functional form adequately describes the measurements, enabling the determination of the parameters \(a\), \(b\), and \(c\).

\subsection{\(\tau\) Energy Dependency}

\begin{figure}[H]
    \centering
    \includegraphics[width=\textwidth]{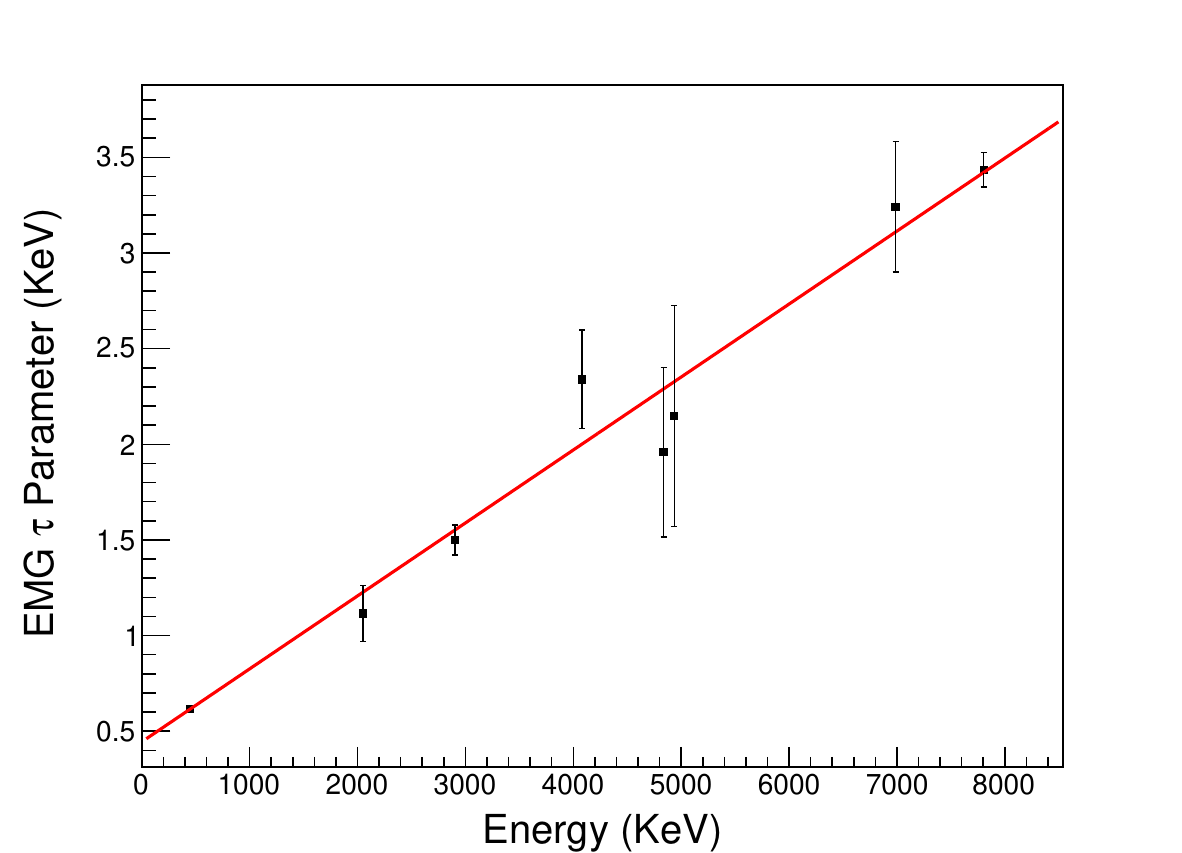}
    \caption{Energy depndency of the EMG \(\tau\) parameter.}
    \label{fig:tau_energy_fit}
\end{figure}

The parameter \(\tau\), which describes the exponential decay component of the EMG fit, was found to have a linear dependency on energy, as shown in Figure \ref{fig:tau_energy_fit}.

\section{Contributions of SE and DE Peaks to Physical Peaks}

When working with \(\gamma\) detectors, if the incident \(\gamma\) energy exceeds 1022 keV, pair production events can occur, resulting in the creation of two 511 keV annihilation \(\gamma\) rays. When only one of these \(\gamma\) rays escapes the detector while the other is fully absorbed, 511 keV are effectively lost, leading to the appearance of a secondary peak in the spectrum at 511 KeV below the original energy of the \(\gamma\) ray, known as the single escape (SE) peak. Conversely, if both annihilation \(\gamma\) rays escape, a double escape (DE) peak arises at 1022 KeV below the original energy of the \(\gamma\) ray.

Examining the complex \(\gamma\) spectrum of \(^{23}\text{Mg}\) and other isotopes, where numerous peaks are present, escape peaks may overlap with actual physical photo peaks from different \(\gamma\) transitions. To assess potential overlaps, the intensity ratios between the photo peak and its SE and DE peaks needed to be determined. Clean peaks, with high certainty of no overlap between escape peaks and physical peaks, were selected for this analysis to establish these intensity ratios.

\begin{figure}[H]
    \centering
    \includegraphics[width=\textwidth]{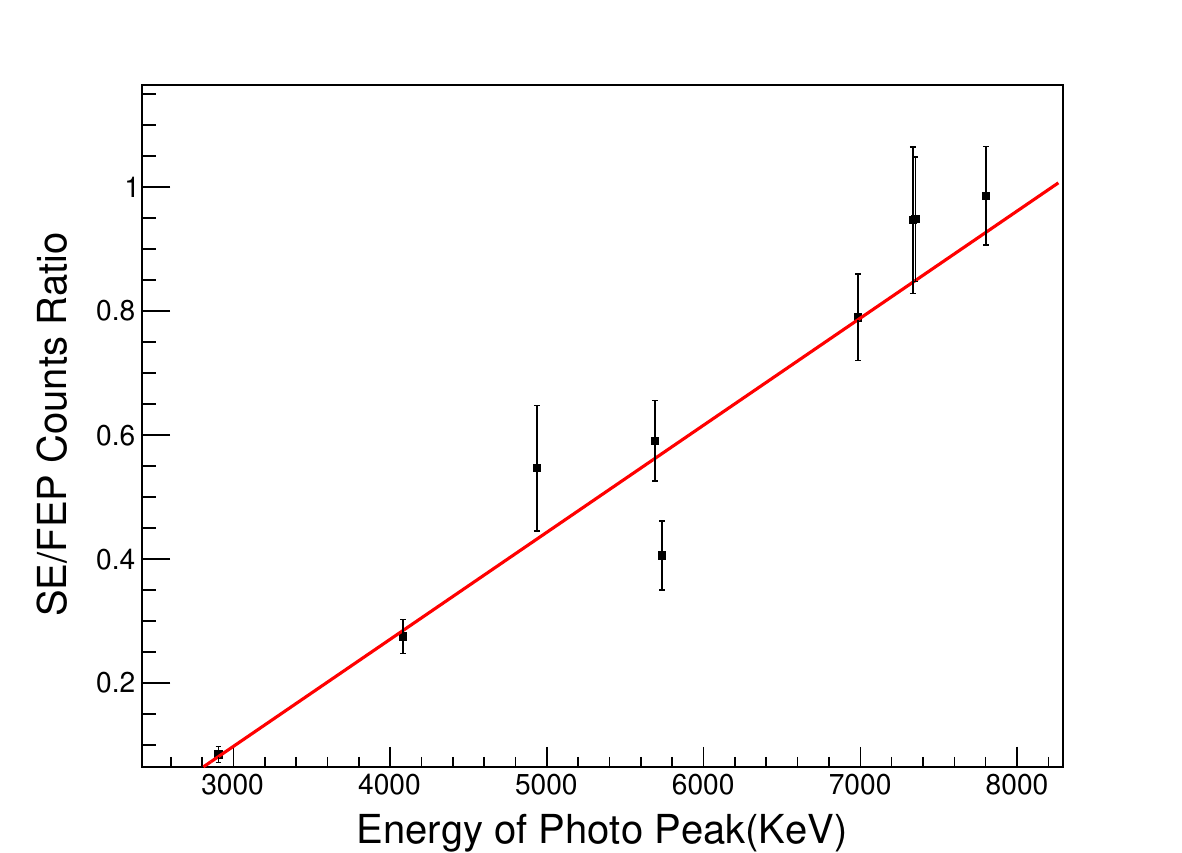}
    \caption{Ratio of SE intensity to FEP intensity as a function of FEP energy, demonstrating its linear relationship due to increasing pair production dominance at higher energies.}
    \label{fig:SE_FEP_ratio}
\end{figure}

\begin{figure}[H]
    \centering
    \includegraphics[width=\textwidth]{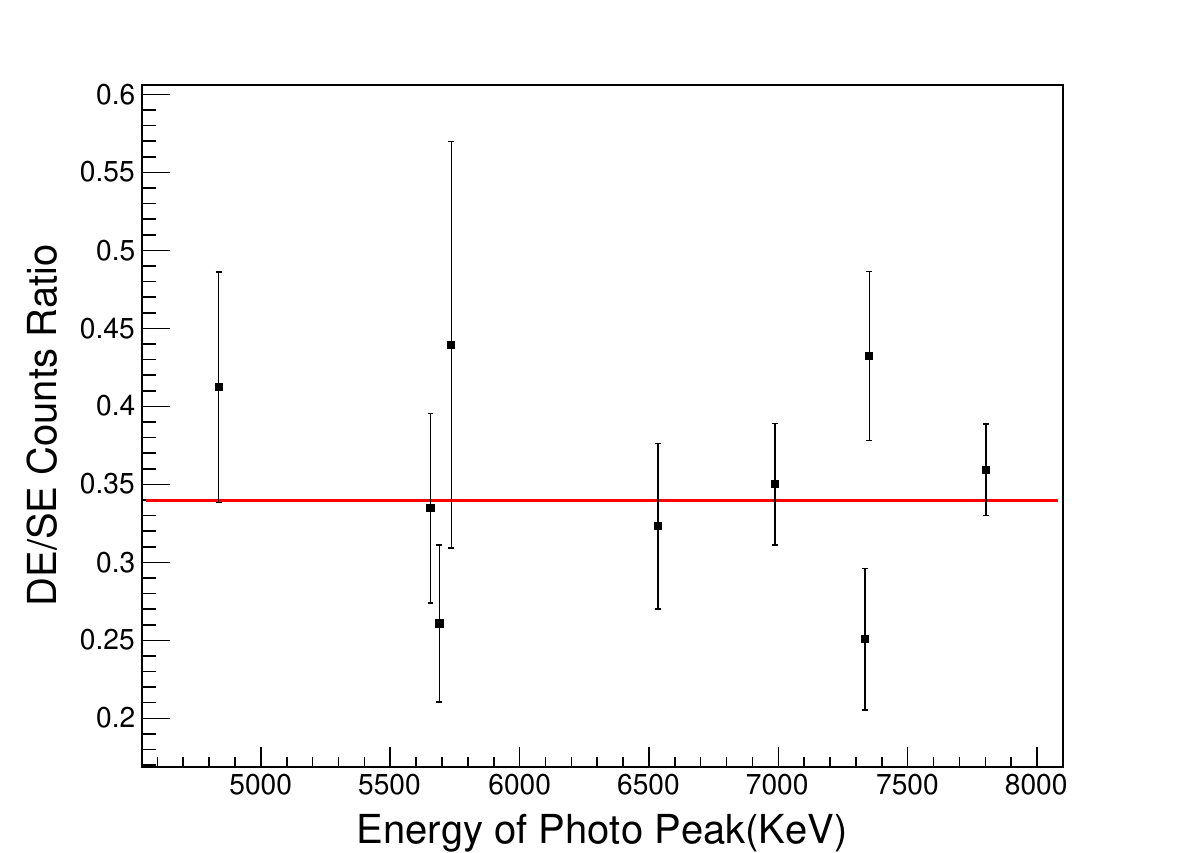}
    \caption{Constancy of the DE intensity to SE intensity ratio, determined to be \(0.339 \pm 0.016\).}
    \label{fig:SE_DE_ratio}
\end{figure}

The ratio of SE intensity to Full Energy Peak (FEP) intensity was found to be linear with the energy of the FEP, as shown in Figure \ref{fig:SE_FEP_ratio}. This linearity arises because pair production becomes increasingly significant as the energy of the incident \(\gamma\) rays increases. Figure \ref{fig:SE_DE_ratio} shows that the DE intensity to SE intensity ratio was determined to be a constant value of \(0.339 \pm 0.016\).

By incorporating these relationships, the contributions of escape peaks to physical peaks could be more accurately accounted for in the analysis. This was achieved by adding a fixed peak, either at FEP-511 keV or FEP-1022 keV, with the appropriate escape peak intensity to the physical peak fitting process. This accounts for both the intensity and mean shifts in overlapping physical peaks.

\begin{figure}[H]
    \centering
    \includegraphics[width=\textwidth]{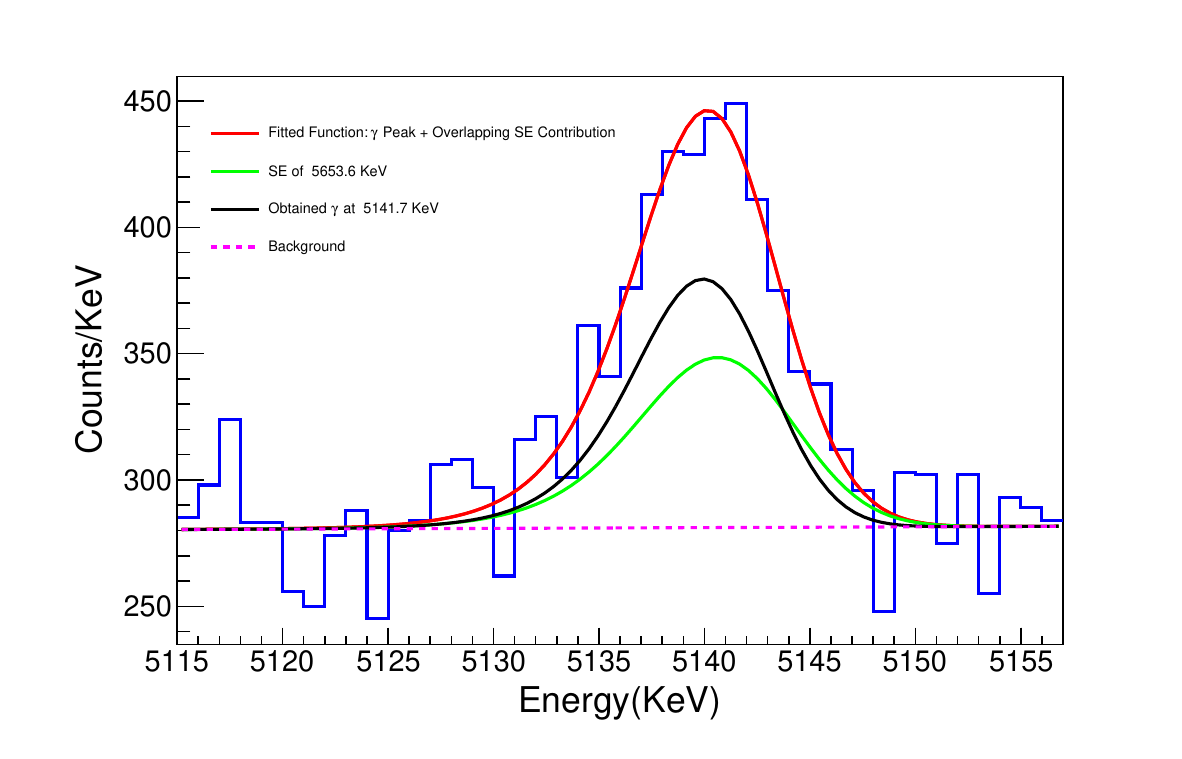}
    \caption{Example of double-peak fitting showing the overlap of a 5142 keV physical peak with a 5654 keV SE peak at 5143 KeV, highlighting the significant impact on fitted physical peak intensity and mean.}
    \label{fig:SE_contribution_example}
\end{figure}

An example of this approach is shown in Figure \ref{fig:SE_contribution_example}, which illustrates a double-peak fitting to physical peaks at 5142 keV and 5655 keV, both originating from \(^{23}\text{Mg}\). The 5142 keV peak overlaps with the SE of the 5655 keV peak. The 5655 keV SE appears at 5144 keV, affecting the perceived intensity and position of what might originally seem to be a single peak around 5143 keV. This example highlights the importance of correctly accounting for SE and DE contributions, which enhances the understanding of the overlapping physical peaks and helps distinguish regular escape peaks from those overlapping with actual physical peaks.

The uncertainties associated with the mean and intensity of the measured peak, which overlaps with the escape peak, are derived from various sources, in addition to the uncertainties related to the fitted parameters. To ensure that the final uncertainties are accurately determined, the uncertainties of both the mean and intensity of the FEP, as well as the uncertainty of the intensity ratio between the FEP and the escape peak, must be propagated. To achieve this, the fitting of the overlapping peaks was conducted multiple times, with the parameters of the escape peak varied within their uncertainty range. This methodology facilitated a thorough assessment of the maximum uncertainty in the parameters of the measured overlapping physical peak.

\section{\(\beta\) Delayed Protons of \(^{23}\text{Al}\)}

The next step toward constructing a complete decay scheme involves detecting and organizing the emitted protons within the decay process. After \(^{23}\text{Al}\) decays into \(^{23}\text{Mg}\), there is a small probability that it will decay further via proton emission into \(^{22}\text{Na}\), provided that the \(^{23}\text{Mg}\) excited state populated has sufficient energy, exceeding the required proton separation energy. Given that for \(^{23}\text{Al}\) the released energy is $Q_\beta$+ = 11199.7 (3) keV , and the proton separation energy (\(S_p\)) of \(^{23}\text{Mg}\) is 7581.25(14) keV \cite{nndc}, a broad range of excited states possess sufficient energy for proton emission.

In this experiment, protons were measured using the Proton Detector component of the GADGET system, which achieves high efficiency and low \(\beta\) background for energies up to about 1 MeV. Beyond this energy, detection efficiency declines rapidly as faster protons exit the active detection volume before depositing all their energy. It is known that \(^{23}\text{Al}\) produces protons at energies both below and above 1 MeV; therefore, the measured protons at lower energies will be reported, alongside adopted data for higher energy protons, to complete the decay scheme.

For the low-energy protons measured, a right-tailed Exponentially Modified Gaussian (EMG) function was fitted, as the peak's wider tail is expected on the higher energy side of the mean. This is due to \(\beta\) summing. The \(\beta\) particles do deposit some small amount of energy in the active volume, and each proton is accompanied by a positron. The fitted function is:

\begin{align}
f(x; N, \mu, \sigma, \tau) = &\; \frac{N}{2\tau} 
\exp\left[\frac{1}{2} \left( \frac{\sigma}{\tau} \right)^2 + \frac{x + \mu}{\tau}\right] \notag \\
&\times \mathrm{erfc}\left[\frac{1}{\sqrt{2}}\left(\frac{\sigma}{\tau} + \frac{x + \mu}{\sigma}\right)\right]
\end{align}

The energy calibration of the proton spectrum is based on proton peak energies reported by Sallaska et al. \cite{Sallaska2010} for values up to 583 keV, and by Saastamoinen et al. \cite{Saastamoinen2011} for the 866-keV proton peak energy. The different pads were calibrated to these known values and gain-matched accordingly.

Regarding detection efficiency, while the proton detector efficiency approximates unity, some minor losses occur due to protons being absorbed by the anode and cathode planes, or high-energy protons escaping the active detection volume prematurely. The efficiency was calculated using a GEANT4 simulation \cite{Agostinelli2003}, which used the same beam spatial distribution described in the SeGA efficiency simulation. Efficiency values were assigned a conservative relative systematic uncertainty of 3\%, derived from variations in beam distribution values.

\section{Half-life of \(^{23}\text{Al}\)}

To determine the half-life of \(^{23}\text{Al}\), our analysis utilized the timing of each \(\gamma\) detection recorded in addition to its energy. The experimental setup was operated in a pulsed mode, where the beam was active for 0.5 seconds and subsequently turned off for another 0.5 seconds, allowing for measurement of decay processes. By recording \(\gamma\) decay events per unit time, a histogram was constructed for the number of measured \(\gamma\) rays as a function of time elapsed since the start of each beam cycle, summing across all cycles of the experiment.

The histogram was fitted with two exponential functions, both sharing the same half-life parameter, reflecting the fact that isotope buildup and decay in the detector exhibit identical half-life characteristics. To isolate the half-life of \(^{23}\text{Al}\) from background and contamination contributions, the histogram was generated specifically for the 451 keV gamma peak, the most intense peak from \(^{23}\text{Al}\) via its daughter \(^{23}\text{Mg}\). This counts-per-time histogram included gammas with energies in the range 450 \(\pm\) 3 keV.

Noise contributions were mitigated by constructing an additional histogram using the same energy width of 6 keV but centered around background energies, specifically 443-446 keV and 454-457 keV, to estimate noise effects. The noise histogram was subsequently subtracted from the 451 keV gated histogram, yielding a cleaned counts-versus-time dataset primarily influenced by non-noise \(\gamma\) events.

The following mathematical function was applied to the histogram data:

\begin{equation}
f(t) = 
\begin{cases} 
B + A_\text{on} \cdot \left[1 -  e^{-\frac{\ln(2) \cdot (t - t_\text{on})}{T_{1/2}}}\right], & \text{for } 50 \text{ ms} < t < 490 \text{ ms, (beam-on phase)} \\ 
B + A_\text{off} \cdot e^{-\frac{\ln(2) \cdot (t - t_\text{off})}{T_{1/2}}}, & \text{for } 550 \text{ ms} < t < 990 \text{ ms, (beam-off phase)}
\end{cases}
\end{equation}

where \(B\) is a constant background parameter, \(A_\text{on}\) and \(A_\text{off}\) are amplitude parameters, \(t_\text{on}\) and \(t_\text{off}\) are timing offsets for the onset of each beam state, and \(T_{1/2}\) is the half-life to be determined. Separate fits were conducted for the beam-on and beam-off time ranges while using the same background and half-life parameters. Measurements from the first 50 ms and last 10 ms of each beam state, characterized by fluctuations and instability, were not considered.

\begin{figure}
    \centering
    \includegraphics[width=\textwidth]{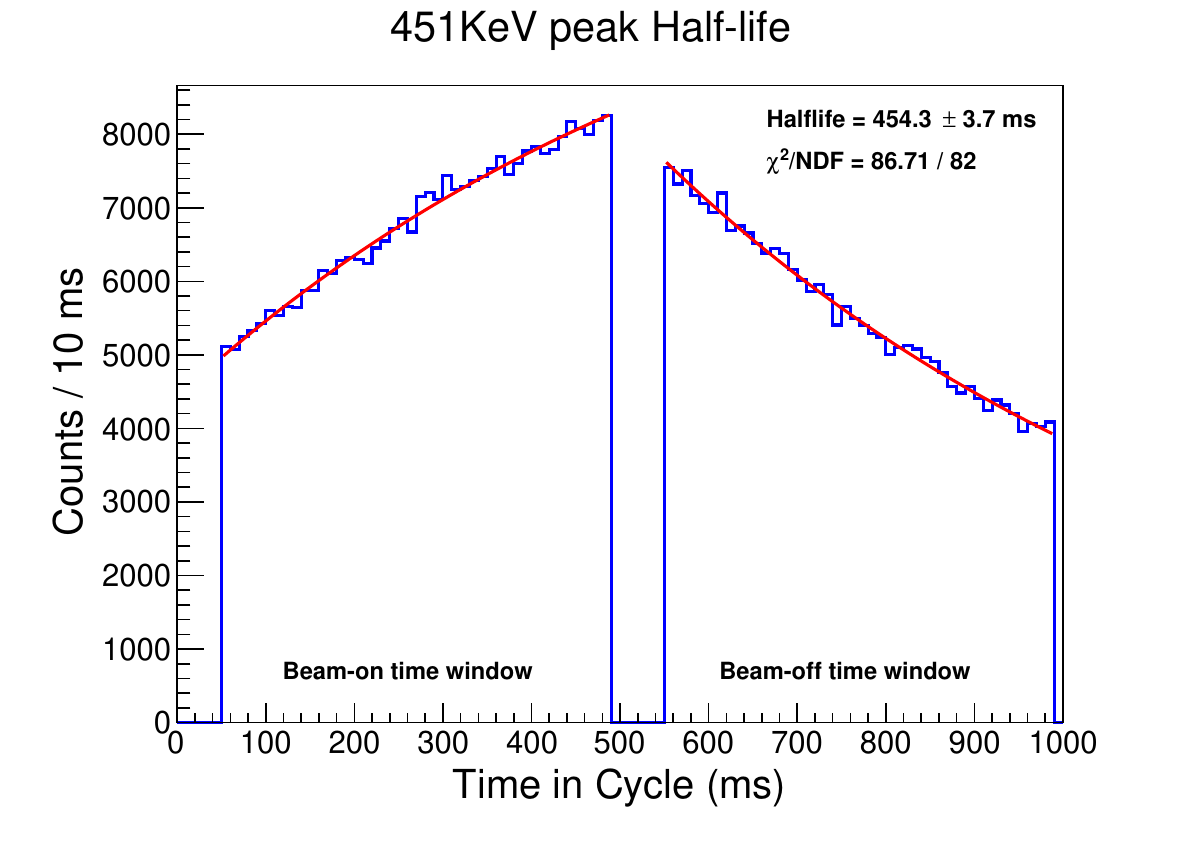}
    \caption{Fitted curve for the half-life determination of \(^{23}\text{Al}\) using \(\gamma\) decay timing data.}
    \label{fig:al23_halflife_curve}
\end{figure}

\begin{figure}
    \centering
    \includegraphics[width=\textwidth]{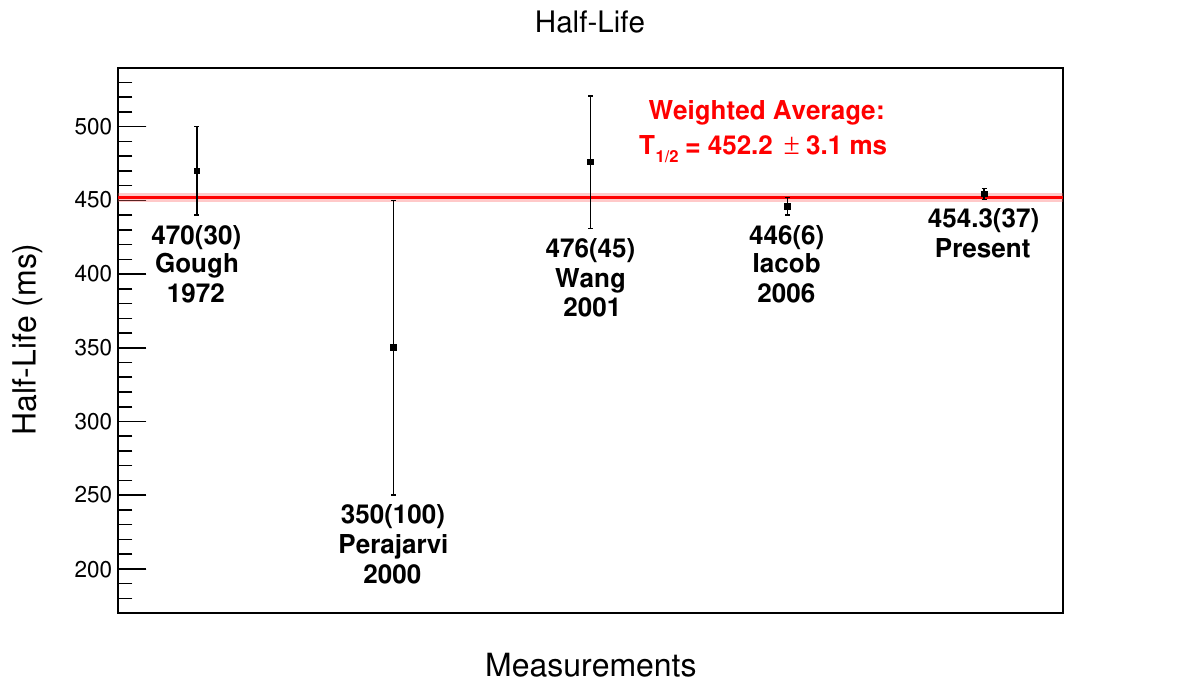}
    \caption{Measurement of the half-life of \textsuperscript{23}Al. The results are compared with previously measured values from Gough et al. \cite{Gough1972}, Peräjärvi \cite{Perajarvi2000}, Wang et al. \cite{Wang2001} and \cite{Wang2002}, and Iacob et al. \cite{Iacob2006}.}

    \label{fig:al23_halflife}
\end{figure}

The fit result is displayed in Figure \ref{fig:al23_halflife_curve}, yielding a half-life of \(454.3 \pm 3.7\) ms for \(^{23}\text{Al}\). This measurement agrees with previously reported values and represents the most precise determination to date, as shown in Figure \ref{fig:al23_halflife}.

%% file: 7.conclusion.tex
\chapter{Results and Discussion}
\label{chap:conclusion}

\section{\(\beta\) Delayed Gammas of \(^{23}\text{Al}\)}

Following the comprehensive analyses presented in previous sections, the list of \(\beta\)-delayed \(\gamma\) rays of \(^{23}\text{Mg}\) measured in this experiment is hereby detailed. A total of 48 \(\gamma\) transitions were identified as originating from \(^{23}\text{Mg}\), with 19 observed for the first time. Notably, no novel excited energy levels were discovered through these observations; rather, new decay pathways were documented.

Peaks originating from the \(^{23}\text{Al}\) beam were identified by excluding background peaks, removing escape peaks, and discounting peaks matching known contaminant transitions. Once these peaks were accounted for, the \(\gamma\)-ray transitions specific to \(^{23}\text{Mg}\) were organized.

Initially, the measured \(\gamma\) rays were compared against the known \(^{23}\text{Mg}\) transitions, facilitating the identification of previously documented transitions. Subsequently, unidentified \(\gamma\) rays were subjected to further analysis. Energy differences corresponding to potential decay paths between established initial and final excited states—confirmed to be populated from other observed known \(\gamma\) transitions—were scrutinized. Consequently, all new \(\gamma\) transitions were matched to known initial and final excited states.

Where sufficient intensity was observed, these transitions were verified via \(\gamma\)-\(\gamma\) coincidence analysis, as presented in Table \ref{tab:gamma_coincidences}. However, some newly identified \(\gamma\) rays, although statistically significant, had inherently low count statistics. This low count rate made it unreasonable to expect an observable \(\gamma\)-\(\gamma\) coincidence due to the reduced efficiency involved in it. Nonetheless, transitions were assigned based on energy alignment with known states, as this provided the most straightforward and consistent interpretation.

Ultimately, all detected gammas conformed to possible transitions derived from established excited states, leading to the conclusion that no new excited states were observed. The results were collated into the comprehensive table of \(\beta\)-delayed gammas for \(^{23}\text{Al}\), presented in Table \ref{tab:beta_delayed_gammas}.

\renewcommand{\arraystretch}{1.5}
\begin{longtable}{ccccc}
    \caption{Measured \(\beta\)-delayed \(\gamma\) rays of \(^{23}\text{Al}\) through \(^{23}\text{Mg}\). Intensities are relative to the intensity of the 450.7 keV \(\gamma\)-ray. \(\gamma\) rays marked with an asterisk (*) are observed for the first time. The recoil energy correction was incorporated into the calculation of the excited state energies.} \\
    \label{tab:beta_delayed_gammas} \\
    \hline
    \(E_{\gamma}\) (keV) & \(I_{\gamma}\) (relative \%) & \(E_i\) (keV) & \(E_f\) (keV) \\
    \hline
    \endfirsthead
    \caption[]{(continued)} \\
    \hline
    \(E_{\gamma}\) (keV) & \(I_{\gamma}\) (relative \%) & \(E_i\) (keV) & \(E_f\) (keV) \\
    \hline
    \endhead
    450.75 (23) & 100 & 451 & 0 \\
    662.97 (20) & 0.281 (49) & 2714 & 2051 \\
    1599.50 (12) & 12.63 (82) & 2051 & 451 \\
    1831.58 (58) & 0.131 (34) & 5692 & 3860 \\
    1906.72 (28) & 0.270 (39) & 2358 & 451 \\
    1966.94 (50) & 0.180 (35) & 4681 & 2714 \\
    2050.77 (14) & 1.80 (12) & 2051 & 0 \\
    2263.26 (24) & 0.492 (51) & 2714 & 451 \\
    2358.34 (88) & 0.126 (42) & 2358 & 0 \\
    2453.52 (17) & 2.62 (18) & 2904 & 451 \\
    2630.82 (30) & 0.292 (33) & 4681 & 2051 \\
    2685.32 (43)* & 0.150 (26) & 6545 & 3860 \\
    2769.68 (58) & 0.114 (25) & 2770 & 0 \\
    2904.21 (21) & 3.90 (26) & 2904 & 0 \\
    3047.21 (65)* & 0.091 (24) & 6907 & 3860 \\
    3174.37 (34)* & 0.109 (25) & 7856 & 4680 \\
    3238.19 (33) & 0.449 (41) & 5288 & 2051 \\
    3415.20 (73) & 0.131 (28) & 6130 & 2714 \\
    3640.89 (59)* & 0.227 (33) & 5692 & 2051 \\
    3669.86 (57)* & 0.240 (35) & 6574 & 2904 \\
    3859.59 (40)* & 0.609 (53) & 3860 & 0 \\
    4082.40 (40)* & 1.71 (12) & 6987 & 2904 \\
    4187.80 (10) & 0.127 (34) & 6239 & 2051 \\
    4230.17 (45) & 0.699 (60) & 4681 & 451 \\
    4465.60 (14)* & 0.187 (37) & 6515 & 2051 \\
    4836.41 (51) & 1.222 (95) & 5288 & 451 \\
    4898.10 (11)* & 0.205 (45) & 7803 & 2904 \\
    4935.88 (56)* & 0.970 (81) & 6987 & 2051 \\
    5141.69 (81) & 0.645 (94) & 7856 & 2714 \\
    5242.10 (15) & 0.620 (14) & 5692 & 451 \\
    5260.90 (11)* & 0.185 (61) & 5711 & 451 \\
    5289.30 (11)* & 0.256 (58) & 5288 & 0 \\
    5653.63 (73) & 0.890 (85) & 5654 & 0 \\
    5689.68 (65) & 2.03 (15) & 5692 & 0 \\
    5709.00 (11)* & 0.248 (61) & 5711 & 0 \\
    5735.81 (60) & 1.78 (14) & 7787 & 2051 \\
    5751.71 (67) & 1.24 (11) & 7803 & 2051 \\
    6062.29 (75) & 4.84 (38) & 6515 & 451 \\
    6122.36 (67) & 4.68 (34) & 6574 & 451 \\
    6513.00 (12)* & 0.293 (73) & 6515 & 0 \\
    6534.76 (77)* & 2.42 (20) & 6987 & 451 \\
    6542.97 (95)* & 1.02 (14) & 6545 & 0 \\
    6573.34 (75)* & 3.47 (24) & 6574 & 0 \\
    6905.70 (10)* & 0.77 (10) & 6907 & 0 \\
    6985.74 (86)* & 4.90 (34) & 6987 & 0 \\
    7335.09 (92) & 7.16 (48) & 7787 & 451 \\
    7351.23 (92) & 10.75 (71) & 7803 & 451 \\
    7802.46 (92) & 21.30 (14) & 7803 & 0 \\
    \hline
\end{longtable}

\subsection{\(^{22}\text{Na}\) \(\gamma\)-Rays}

In addition to the \(\gamma\) rays presented in the \(^{23}\text{Mg}\) table, two additional \(\gamma\) rays were measured that warrant separate discussion. The first is a \(\gamma\) ray at 583 keV, previously reported by Friedman \cite{Friedman2020} based on the data of this experiment. This \(\gamma\) ray originates from the transition of \(^{22}\text{Na}\) from its first excited state to the ground state. The population of this first excited state is facilitated by a proton emitted from the 8762(6) keV excited state of \(^{23}\text{Mg}\), which is consistent with the known proton separation energy of \(S_p = 7581.25(14)\) keV \cite{nndc}. 

\begin{figure}[H]
    \centering
    \includegraphics[width=\textwidth]{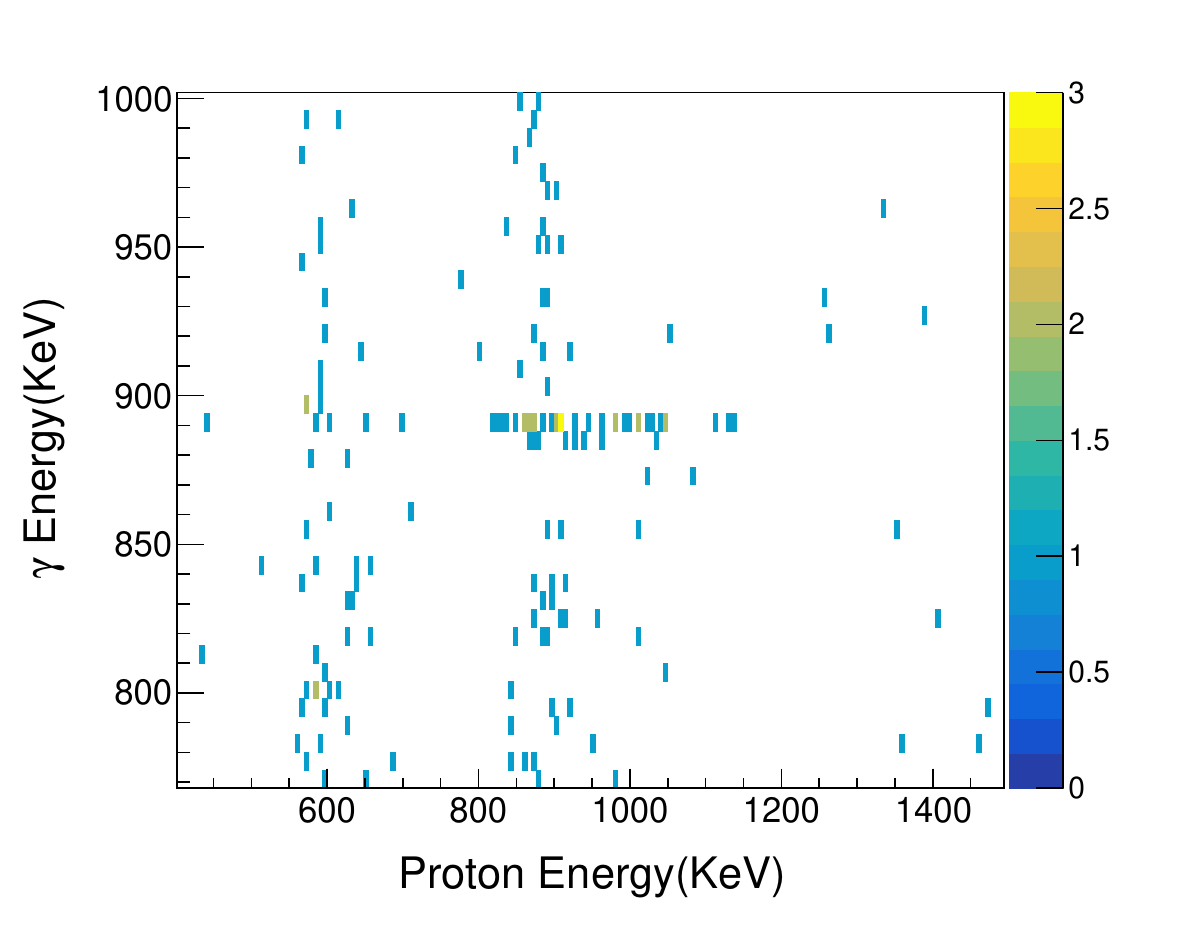}
    \caption{Zoom-in on the 2D \(\gamma\)-proton coincidence spectrum, higlighting a coincidence between a 890 keV \(\gamma\) and a 898 KeV proton.}
    \label{fig:890_p-g}
\end{figure}

Another significant \(\gamma\) ray observed was at 890 keV, which did not correspond to any possible transitions of \(^{23}\text{Mg}\). Upon further examination, it was revealed that this \(\gamma\) ray is in coincidence with a proton of approximately 890 keV. This observation is illustrated in Figure \ref{fig:890_p-g}. A process of 2D noise reduction from the proton-\(\gamma\) coincidence histogram was used to confirm a significant event post-noise reduction.

The coincidence suggests that the 890 keV \(\gamma\) ray also results from a transition in \(^{22}\text{Na}\), following proton emission from \(^{23}\text{Mg}\). Interestingly, the 890 keV \(\gamma\) is a recognized transition from the third excited state of \(^{22}\text{Na}\), confirming the discovery of a new proton-\(\gamma\) transition previously unobserved. These findings directly underscore the capabilities of the GADGET system, which was expressly designed to detect proton-\(\gamma\) coincidences. This system’s ability to measure both proton and \(\gamma\) events concurrently with high certainty exemplifies its advantage.

The discovered proton associated with the 890 keV \(^{22}\text{Na}\) \(\gamma\) will be further analyzed in a subsequent section focusing on proton measurements.

\section{\(\beta\) Delayed Protons of \(^{23}\text{Al}\)}

A total of six protons were measured in this experiment. These protons are detailed in Table \ref{tab:proton_data}, which summarizes the energies, relative intensities, and transitions involved.

To measure the intensity of the proton peaks, the combined spectrum of pads A-E was utilized, providing maximal efficiency. According to simulations, the detection efficiencies at energies of 204, 275, 583, and 866 keV were 0.98, 0.98, 0.97, and 0.93, respectively. The higher efficiency of this combined spectrum comes at the expense of resolution, which limits the ability to distinguish between proximate protons at 583-595 keV and 866-898 keV. To accurately determine the means of these double-peak protons and measure the intensity ratio between each pair, the spectrum of pad A alone, offering superior resolution, was also fitted.

\begin{figure}[H]
    \centering
    \includegraphics[width=\textwidth]{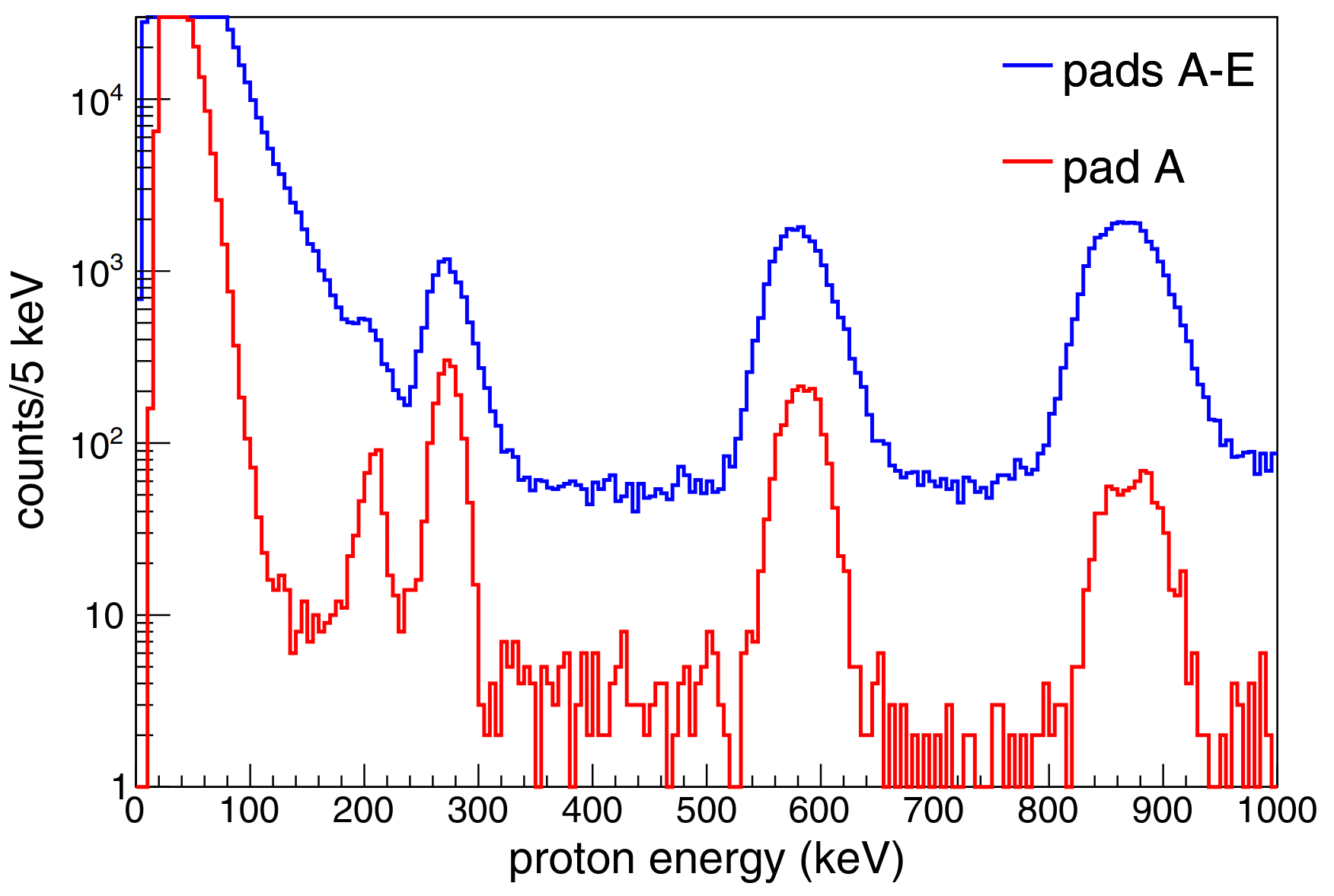}
    \caption{Proton spectra for pads A-E (in blue) and the central pad (in red). The data illustrate the superior efficiency achieved with multiple pads compared to the enhanced resolution obtained when using pad A alone. Adopted from Friedman et al. \cite{Friedman2020}.}
    \label{fig:proton_peak_means}
\end{figure}

In the higher-resolution pad A spectrum, each double peak was fitted with a double EMG function, with both peaks sharing the same \(\sigma\) and \(\tau\) parameters. This fitting revealed significant separation between means in alignment with expected values for the proton energies. Subsequently, the actual intensities of the peaks were derived by taking the intensity results from the combined A-E spectrum fit and apportioning them according to the ratio obtained from the pad A spectrum fit. 

It should be noted that for the 204 and 275 keV peaks, a linear background model did not suffice; instead, an exponentially decaying model was employed to adequately describe the background.

The protons at 595 and 898 keV, which faced challenges due to resolution limitations when separating from neighboring peaks at 583 and 866 keV, were observed for the first time in this experiment. The GADGET system's unique capability of measuring proton-\(\gamma\) coincidences allowed for the distinction of these new protons, which populate two excited levels of \(^{22}\text{Na}\). In both instances, the observed \(\gamma\)-proton coincidence was not intense enough to account for the total measured intensity, necessitating the conclusion of a double-peak structure for protons in each case. This conclusion finds support in the higher-resolution fits to the pad A spectrum. The double-peak fit near the 866 keV proton indicates that the mean energy of the second peak is 898 keV. By summing this with the \(S_p\) of \(^{23}\text{Mg}\) and the 890 keV excited state of \(^{22}\text{Na}\) that it populates, an approximate energy level of 9369 keV in \(^{23}\text{Mg}\) is suggested. Given the existence of a known excited level of \(^{23}\text{Mg}\) at 9374(8)KeV \cite{nndc}, it is concluded that the newly detected proton at 898 keV originates from this known level.

\begin{table}[H]
    \centering
    \caption{Measured proton energies and relative intensities. Intensities are relative to the combined intensity of the 866 and 898 keV protons, consistent with prior publications.}
    \label{tab:proton_data}
    \begin{tabular}{cccc}
        \hline
        \(E_{\text{c.m.}}\) (keV) & \(I\) (relative) & \(^{23}\text{Mg}\) \(E_i\) (keV) & \(^{22}\text{Na}\) \(E_f\) (keV) \\
        \hline
        204 & 0.0623(79) & 7787 & 0 \\
        275 & 0.291(17) & 7856 & 0 \\
        583 & 0.189(13) & 8165 & 0 \\
        595 & 0.498(29) & 8762 & 583 \\
        866 & 0.687(47) & 8448 & 0 \\
        898 & 0.313(32) & 9374 & 890 \\
        \hline
    \end{tabular}
\end{table}

\subsection{Adopted Protons}

To complete the full decay scheme, it is necessary to consider previously measured protons with energies exceeding 1 MeV, which were beyond the detection capabilities of our system. These protons are crucial for calculating the \(\beta\) feeding ratios from \(^{23}\text{Al}\) into the excited states of \(^{23}\text{Mg}\). Protons with energies higher than 1 MeV were adopted from the work by Kirsebom et al. \cite{Kirsebom2011}. The reported intensities were relative to the 866 keV peak, which, as clarified by our results, encompass the combined intensities of the 866 and 898 keV protons. All adopted protons populate the ground state of \(^{22}\text{Na}\).

The range in which Kirsebom's experiment exhibited good detection capabilities was primarily at energies above 500 keV. Consequently, our results were not directly comparable at lower energies; however, both experiments maintained reliable detection at 866 keV. Therefore, reporting intensities relative to this reference point remains valid for both sets of results. The adopted protons are presented in Table \ref{tab:adopted_protons}.

\begin{table}[H]
    \centering
    \caption{Adopted protons from Kirsebom et al. \cite{Kirsebom2011}. Proton energies are calculated from the reported \(E_i\) values of \(^{23}\text{Mg}\) and the known \(S_p\). Intensities are relative to the combined 866 KeV and 898 keV double peak intensity.}
    \label{tab:adopted_protons}
    \begin{tabular}{ccc}
        \hline
        Proton Energy (keV) & \(E_i\) (\(^{23}\text{Mg}\)) (keV) & Relative Intensity \\
        \hline
        866 & 8448 & 1 \\
        998 & 8579 & 0.0155 (10) \\
        1200 & 8781 & 0.0317 (15) \\
        1259 & 8840 & 0.0142 (10) \\
        1324 & 8905 & 0.0411 (16) \\
        1442 & 9023 & 0.0077 (7) \\
        1521 & 9102 & 0.0078 (14) \\
        1563 & 9144 & 0.042 (20) \\
        1740 & 9321 & 0.0144 (10) \\
        1841 & 9422 & 0.0521 (19) \\
        1888 & 9469 & 0.0204 (13) \\
        2023 & 9604 & 0.0062 (7) \\
        2101 & 9682 & 0.0020 (5) \\
        \hline
    \end{tabular}
\end{table}

As mentioned in the Experimental Setup chapter, protons in our experiment were measured only during intervals of 0.5 seconds when the beam was turned off. To obtain an accurate absolute intensity of the protons for comparison with the gamma intensities, it is essential to account for dead time. Consequently, the intensities of the \( \beta \)-delayed protons at 866 keV and 898 keV were normalized so that their combined BR equals 0.41(1)\%, as reported by Saastamoinen et al. \cite{Saastamoinen2011}. All other proton intensities, for which we have already established values relative to the sum of the 866 keV and 898 keV intensities, were then normalized accordingly.

\section{Decay Scheme of \(^{23}\text{Al}\)}

This section presents the complete decay scheme of \(^{23}\text{Al}\), excluding follow-up \(\beta\) decays from \(^{23}\text{Mg}\) and \(^{22}\text{Na}\).
 The scheme is constructed using all previously presented \(\gamma\) and proton detections, alongside adopted data. Known energy levels ans Spin-parity values were adopted from the National Nuclear Data Center (NNDC) \cite{nndc}.

The BR of the \(\beta\) feeding for each \(^{23}\text{Mg}\) populated level is calculated as follows: 
\begin{equation}
\beta_\text{feeding} = \gamma_\text{emission} + P_\text{emission} - \gamma_\text{feeding}
\end{equation}
Here, \(\gamma\)-emission represents the \(\gamma\) counts detected from the level, \(P\)-emission represents the proton emission from the level, and \(\gamma\)-feeding accounts for the measured \(\gamma\) counts populating the level from \(\gamma\) decays from higher excited states. All counts are corrected for detection efficiencies.

For the ground state of \textsuperscript{23}Mg, no \( \gamma \) emission data was available, necessitating the adoption of alternative evaluation methods. Consequently, the experimental result from Iacob et al. \cite{Iacob2006}, which provides a \( \beta \) feeding BR of \( 36.3(16)\% \) from \textsuperscript{23}Al to the ground state of \textsuperscript{23}Mg, is utilized in the calculations of this work. All other BR values are expressed as percentages, with this value taken into account as a baseline: the \( 36.3\% \) corresponds to a decay count representing \( 36.3\% \) of the total, inclusive of all measured decays to excited states.

\begin{figure}
    \centering
    \hspace*{-2.6cm}
    \includegraphics[width=1.35\textwidth]{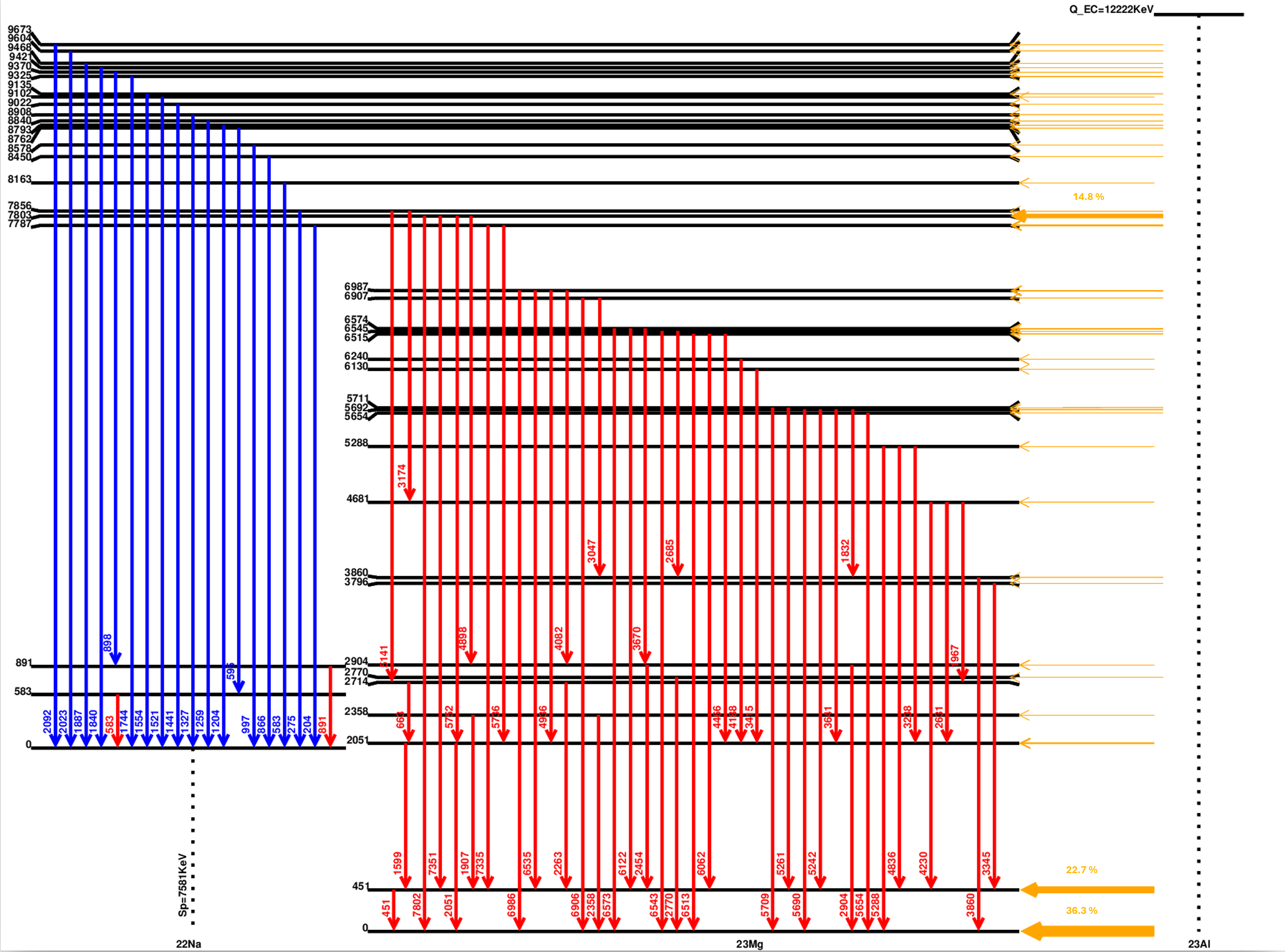}
    \caption{Schematic representation of the full decay scheme of \(^{23}\text{Al}\), excluding further \(\beta\) decays from \(^{23}\text{Mg}\) and \(^{22}\text{Na}\). The \(\beta\) decays are represented by orange arrows, with the arrow width indicating the \(\beta\)-feeding ratio. \(\gamma\) transitions are shown as red arrows, while proton transitions are depicted in blue.}
    \label{fig:full_decay_scheme}
\end{figure}

The \(\beta\) feeding values were essential for calculating the log-\(ft\) values using the NNDC log-\(ft\) calculator \cite{nndc_logft}. This detailed decay scheme is encapsulated in Table \ref{tab:energy_levels_branching_ratios}, summarizing energy levels, \(\beta\) feeding percentages, spin-parity values, and log-\(ft\) values. The complete decay scheme is visually depicted in Figure \ref{fig:full_decay_scheme}.

\renewcommand{\arraystretch}{1.5}
\begin{longtable}{>{\centering\arraybackslash}m{3cm} >{\centering\arraybackslash}m{3cm} >{\centering\arraybackslash}m{3cm} >{\centering\arraybackslash}m{3cm} >{\centering\arraybackslash}m{3cm}}
    \caption{Energy levels of \(^{23}\text{Mg}\) populated by the \(\beta\) decay of \(^{23}\text{Al}\), along with the corresponding branching ratios and Log-\textit{ft} values. Transitions marked with an asterisk (*) are considered tentative and should be further investigated, as discussed in Section \ref{sec:forbidden_decays}.} \label{tab:energy_levels_branching_ratios} \\
    \hline
    Previously Known Energy Level (KeV) & Measured Energy Level (KeV) & \(\beta\)-Feeding to Level \% & Previously Known \(J^{\pi}\) & Log-\textit{ft} \\
    \hline
    \endfirsthead 
    \caption[]{(continued)} \\ 
    \hline
    Previously Known Energy Level (KeV) & Measured Energy Level (KeV) & \(\beta\)-Feeding to Level \% & Previously Known \(J^{\pi}\) & Log-\(f_t\) \\
    \hline
    \endhead 
    \hline
    \endfoot 
    \hline
    \endlastfoot 
        0 & 0 & 36.3 (16) & \(3/2+\) & 5.309 (14) \\
        450.70 (15) & 450.75 (23) & 22.7 (25) & \(5/2+\) & 5.43 \(\left(^{+6} _{-5}\right)\) \\
        2051.6 (4) & 2050.75 (12) & 3.93 (45) & \(7/2+\) & 5.85 \(\left(^{+6} _{-5}\right)\) \\
        2357.0 (7) & 2357.69 (34) & 0.175 (35)* & \(1/2+\) & 7.13 \(\left(^{+10} _{-8}\right)\) \\
        2714.5 (5) & 2713.96 (26) & 0.000 (81) \footnotemark & \(9/2+\) & - - - \\
        2771.2 (7) & 2769.87 (58) & 0.051 (11)* & \(1/2-\) & 9.81 \(\left(^{+11} _{-9}\right)\) \\
        2905.2 (7) & 2904.41 (13) & 1.93 (20) & \(3/2+\) & 5.96 (5) \\
        3860.6 (7) & 3859.93 (40) & 0.105 (43) & \(3/2+, 5/2+\) & 6.97 \(\left(^{+23} _{-15}\right)\) \\
        4681.5 (7) & 4681.34 (24) & 0.471 (56) & \(7/2+\) & 6.08 \(\left(^{+6} _{-5}\right)\) \\
        5287.5 (8) & 5288.43 (44) & 0.854 (82) & \(3/2+, 5/2+\) & 5.62 \(\left(^{+4} _{-5}\right)\) \\
        5658 (4) & 5654.38 (73) & 0.394 (36) & \(5/2+\) & 5.83 (4) \\
        5690.7 (6) & 5691.53 (38) & 1.34 (15) & \((1/2 \text{ to } 9/2)+\) & 5.28 \(\left(^{+6} _{-5}\right)\) \\
        5712 (8) & 5711.08 (73) & 0.192 (54) & \(1/2+, 3/2+, 5/2+\) & 6.12 \(\left(^{+15} _{-11}\right)\) \\
        6129.3 (7) & 6129.65 (77) & 0.058 (12)* & \(7/2-\) & 6.48 \(\left(^{+11} _{-9}\right)\) \\
        6240.3 (13) & 6239.0 (11) & 0.056 (15)* & \(9/2+\) & 6.45 \(\left(^{+14} _{-11}\right)\) \\
        6513.6 (10) & 6514.97 (57) & 2.36 (21) & \(7/2+\) & 4.71 \(\left(4\right)\) \\
        6540 (3) & 6544.67 (52) & 0.520 (72) & \(3/2+, 5/2+\) & 5.36 \(\left(^{+7} _{-6}\right)\) \\
        6574.6 (10) & 6574.30 (32) & 3.72 (26) & \(5/2+\) & 4.487 \(\left(^{+30} _{-32}\right)\) \\
        6908 (3) & 6907.04 (56) & 0.379 (56) & \(5/2+\) & 5.33 \(\left(^{+7} _{-6}\right)\) \\
        6984 (5) & 6987.00 (36) & 4.43 (31) & \(5/2+\) & 4.226 \(\left(^{+30} _{-32}\right)\) \\
        7784.7 (8) & 7787.21 (52) & 3.99 (25) & \(3/2+, 5/2+\) & 3.853 \(\left(28\right)\) \\
        7803.0 (6) & 7803.29 (40) & 14.85 (94) & \(5/2+\) & 3.274 \(\left(^{+27} _{-29}\right)\) \\
        7855.5 (7) & 7856.15 (33) & 0.454 (52) & \(7/2+\) & 4.76 \(\left(^{+6} _{-5}\right)\) \\
        8163.1 (12) & - - - & 0.0781 (59) & \(5/2+\) & 5.334 \(\left(^{+32} _{-34}\right)\) \\
        8449 (2) & - - - & 0.285 (20) & \(3/2+, 5/2+\) & 4.581 \(\left(^{+30} _{-32}\right)\) \\
        8578 (2) & - - - & 0.00642 (44) & \(3/2+, 5/2+, 7/2+\) & 6.136 \(\left(^{+29} _{-31}\right)\) \\
        8762 (6) & - - - & 0.207 (13) & - - - & 4.488 \(\left(^{+27} _{-29}\right)\) \\
        8793 (8) & - - - & 0.01313 (69) & \(7/2+\) & 5.661 (4) \\
        8840 (3) & - - - & 0.00589 (44) & \(3/2+, 5/2+, 7/2+\) & 5.972 \(\left(^{+32} _{-34}\right)\) \\
        8908 (3) & - - - & 0.01703 (78) & \(5/2+\) & 5.454 (20) \\
        9024 (5) & - - - & 0.00320 (30) & \(3/2+, 5/2+, 7/2+\) & 6.09 (4) \\
        9102 (6) & - - - & 0.00323 (58) & \(3/2+, 5/2+, 7/2+\) & 6.01 \(\left(^{+9} _{-8}\right)\) \\
        9135 (6) & - - - & 0.01740 (93) & \(3/2+, 5/2+, 7/2+\) & 5.247 (23) \\
        9325 (5) & - - - & 0.00597 (44) & \(3/2+, 5/2+, 7/2+\) & 5.531 \(\left(^{+31} _{-34}\right)\) \\
        9374 (8) & - - - & 0.130 (14) & - - - & 4.15 (5) \\
        9421 (4) & - - - & 0.02159 (94) & \(3/2+, 5/2+, 7/2+\) & 4.874 (19) \\
                9468 (5) & - - - & 0.00845 (57) & \(3/2+, 5/2+, 7/2+\) & 5.232 \(\left(^{+29} _{-31}\right)\) \\
        9604 (5) & - - - & 0.00257 (30) & \(3/2+, 5/2+, 7/2+\) & 5.60 \(\left(^{+6} _{-5}\right)\) \\
        9673 (7) & - - - & 0.00083 (21) & \(3/2-\) & 6.01 \(\left(^{+13} _{-10}\right)\) \\
        \hline
\end{longtable}
\footnotetext{A $\beta$ decay to this state is forbidden. Although $\gamma$ rays emitted from the 2714 keV state are observed, $\gamma$ emissions at 1967, 3416, and 5142 keV populate this state through $\gamma$ decays. Considering these contributing transitions, it is concluded that this state is not significantly populated by $\beta$ decay.}

The energy states presented in Table \ref{fig:full_decay_scheme} were calculated based on a comprehensive analysis of the emitted \(\gamma\) rays. For each state, the energies of all measured \(\gamma\) rays were considered, and after accounting for recoil energy, the energies of the initial states were derived. The reported energy values represent a weighted average of the results from all \(\gamma\) emissions. These reported values are in agreement with previously known values, and significant improvements in precision have been achieved for certain states, which previously had uncertainties of several keV. For example, the states at 5654 keV, 5711 keV, and 6987 keV have shown marked improvements, as detailed in the table.

\subsection{\label{sec:astro}Proton Branching Ratios}

The \({}^{22}\text{Na}(p,\gamma){}^{23}\text{Mg}\) reaction serves as the primary destruction channel for \({}^{22}\text{Na}\) during a nova. The key protons at 204 keV and 275 keV emitted from the 7787 keV and 7856 keV excited states of \({}^{23}\text{Mg}\), respectively, are critical resonances for this destruction.

In this experiment, we report the proton BR for these two important \({}^{23}\text{Al}\) beta-delayed protons, which act as probes for the resonance strength of the \({}^{22}\text{Na}(p,\gamma)\) reaction. For the 204 keV proton emitted from the 7787 keV state, our measured \(\gamma\) intensity is consistent with the one reported in \cite{Iacob2006}, hence the proton BR reported by Friedman et al. \cite{Friedman2020} is unchanged.

A new \(\gamma\) branch was measured in this experiment at 3174 keV from the 7856 keV state, which affects its proton BR. However, two known \(\gamma\) branches at 5803 keV and 7404 keV from this state were not observed. Upper limits were established for the intensities of these \(\gamma\) lines, with \(I^{\text{max}}_{5803} = 0.61\%\) and \(I^{\text{max}}_{7404} = 0.36\%\) relative to the intensity of the 451 keV \(\gamma\) ray, at a 99.7\% confidence level. These upper limit values are consistent with previously measured intensities \cite{Sallaska2010}, allowing us to assume their continued existence for subsequent calculations. Taking that into account, for the 275 keV proton emitted from the 7856 keV state, a proton BR of \(\Gamma_p / \Gamma = 0.203(20)\) was obtained. No previous value for this BR ratio was published. However, in order to calculate the astrophysical implications the lifetime of the 7856-keV state must be determined.

\subsection{\label{sec:forbidden_decays}Forbidden \(\beta\) decays}

Some forbidden \(\beta\) transitions appear to have been detected during the experiment. The occurrence of these transitions is considered highly improbable, as their probabilities are at least about \(10^{-4}\) of those for allowed transitions, and the statistics obtained in this experiment are insufficient to detect such low-probability events. The measured forbidden decays are into the states at 2358, 2770, 6240, and 6130 keV. Either the adopted spin-parity values for these states are incorrect, or an explanation must be found for their apparent population in this experiment.

A plausible explanation for the seemingly observed forbidden transitions to the 6240 and 2358 keV states involves the 4188 keV \(\gamma\) ray, which has about 13,000 counts. While initially assigned, based on NNDC data \cite{nndc} to the 6240-keV to 2051-keV transition, it may also result from a transition from a 6545 KeV state to the 2358 KeV state. Additionally, the 5290 keV \(\gamma\), observed with $\sim$26,500 counts and initially assigned to the 5288 keV to GS transition, may instead signify a transition from a 7648 keV state to the 2358 keV state. Summing contributions from these transitions yields approximately 39,500 counts, consistent with the observed $\sim$41,000 counts for the 2358 keV state.

Regarding the 2770 keV and 6130 keV states, \(\gamma\) counts of approximately 12,000 and 14,000 were measured, respectively. Several observed \(\gamma\) rays may contribute to feeding these states. These \(\gamma\) rays were initially assigned to other transitions based on previous experiments \cite{nndc}. The \(\gamma\) ray at 2263 keV, which was measured with approximately 51,000 counts, might include a transition from the initial state of 8393 keV to the final state of 6130 keV. Similarly, the \(\gamma\) ray at 5654 keV, measured with around 92,000 counts, may encompass a transition from an initial state of 8427 keV to a final state of 2770 keV. Additionally, the \(\gamma\) ray at 2631 keV, with a recorded intensity of about 30,000 counts, could involve a transition from the initial state of 8762 keV to the final state of 6130 keV. While these suggested initial states are known states of \(^{23}\text{Mg}\), there is currently no evidence confirming their population in our experiment. 

These transitions represent potential pathways for feeding the questioned states via allowed \(\gamma\) decays, thereby addressing the issue of the forbidden \(\beta\) decay. Unfortunately, the tools and statistics available in our experiment are insufficient to verify these hypotheses. Additional research will be required to confirm their validity.

%% file: main.bbl
\begin{thebibliography}{10}

\bibitem{Friedman2019}
M.~Friedman et~al.
\newblock {\em Nucl. Instrum. Method. Phys. Res. Sect. A}, 940:93, 2019.

\bibitem{Gough1972}
R.~A. Gough, R.~G. Sextro, and Joseph Cerny.
\newblock {\em Phys. Rev. Lett.}, 28:510, 1972.

\bibitem{Perajarvi2000}
K.~Peräjärvi et~al.
\newblock {\em Phys. Lett. B}, 492(1-2):1--7, 2000.

\bibitem{Wang2001}
H.-W. Wang et~al.
\newblock {\em High Energy Physics and Nuclear Physics (China)}, 25:971, 2001.

\bibitem{Wang2002}
H.-W. Wang.
\newblock {\em High Energy Physics and Nuclear Physics (China)}, 26:1117, 2002.

\bibitem{Iacob2006}
V.~E. Iacob et~al.
\newblock {\em Phys. Rev. C}, 74:045810, 2006.

\bibitem{Friedman2020}
M.~Friedman et~al.
\newblock {\em Phys. Rev. C}, 101:052802(R), 2020.

\bibitem{Kirsebom2011}
O.~S. Kirsebom et~al.
\newblock {\em Eur. Phys. J. A}, 47:130, 2011.

\bibitem{Basunia2021}
M.~S. Basunia.
\newblock {\em Nuclear Data Sheets}, 171:1, 2021.

\bibitem{Saastamoinen2011}
A.~Saastamoinen et~al.
\newblock {\em Phys. Rev. C}, 83:045808, 2011.

\bibitem{Trache2012}
L.~Trache et~al.
\newblock In {\em J. Phys: Conf. Series}, volume 337, page 012058, 2012.

\bibitem{Zhai2007}
Y.~Zhai.
\newblock PhD thesis, Texas A\&M University, 2007.

\bibitem{Krane1988}
K.~S. Krane.
\newblock {\em Introductory Nuclear Physics}.
\newblock Wiley John $\&$ Sons, New York, 1988.

\bibitem{nndc}
{National Nuclear Data Center}.
\newblock National nuclear data center, 2024.
\newblock Accessed May 2024.

\bibitem{blatt1979theoretical}
J.~M. Blatt and V.~F. Weisskopf.
\newblock Springer, New York, 1979.

\bibitem{Marti2001}
F.~Marti et~al.
\newblock In F.~Marti, editor, {\em Cyclotrons and Their Applications 2001:
  Sixteenth International Conference}, AIP Conf. Proc. No. 600, pages 64--68,
  New York, 2001. AIP.

\bibitem{Stolz2005}
A.~Stolz et~al.
\newblock {\em Nucl. Instrum. Method. Phys. Res. Sect. B}, 241:858, 2005.

\bibitem{Bazin2009}
D.~Bazin et~al.
\newblock {\em Nucl. Instrum. Method. Phys. Res. Sect. A}, 606:314, 2009.

\bibitem{Giomataris2006}
I.~Giomataris et~al.
\newblock {\em Nucl. Instrum. Method. Phys. Res. Sect. A}, 560:405, 2006.

\bibitem{Mueller2001}
W.~Mueller et~al.
\newblock {\em Nucl. Instrum. Method. Phys. Res. Sect. A}, 466:492, 2001.

\bibitem{Chen2017}
J.~Chen.
\newblock Nuclear data sheets for a = 40.
\newblock {\em Nucl. Data Sheets}, 140:1, 2017.

\bibitem{Martin2007}
M.~J. Martin.
\newblock Nuclear data sheets for a = 208.
\newblock {\em Nucl. Data Sheets}, 108:1583, 2007.

\bibitem{Agostinelli2003}
S.~Agostinelli et~al.
\newblock {\em Nucl. Instrum. Method. Phys. Res. Sect. A}, 506:250, 2003.

\bibitem{Sun2021}
L.~J. Sun, M.~Friedman, et~al.
\newblock {\em Phys. Rev. C}, 103:014322, 2021.

\bibitem{Sallaska2010}
A.~L. Sallaska et~al.
\newblock {\em Phys. Rev. Lett.}, 105:152501, 2010.

\bibitem{nndc_logft}
{National Nuclear Data Center}.
\newblock Log-ft $\beta$-decay calculator, 2024.
\newblock National Nuclear Data Center, Accessed May 2024.

\end{thebibliography}
